\documentclass{article}

\usepackage{PRIMEarxiv}

\usepackage[utf8]{inputenc} 
\usepackage[T1]{fontenc}    
\usepackage{hyperref}       
\usepackage{url}            
\usepackage{booktabs}       
\usepackage{amsfonts}       
\usepackage{nicefrac}       
\usepackage{microtype}      
\usepackage{lipsum}
\usepackage{fancyhdr}       
\usepackage{graphicx}       
\graphicspath{{media/}}     

\usepackage{xspace}
\usepackage{subcaption}
\usepackage{amsmath}
\usepackage{bm}
\usepackage{cite}

\newcommand{\astname}{2019~UO$_{14}$\xspace}
\newcommand{\dv}{\Delta V}
\newcommand{\Isp}{I_\mathrm{sp}}

\graphicspath{{figures/}}

\title{Mission Analysis for the First-Ever Saturn Trojan 2019~UO$_{14}$
}

\author{
  Yuki Takao \\
  Department of Aeronautics and Astronautics \\
  Kyushu University \\
  Fukuoka, 819-0395, Japan \\
  \texttt{takao.yuki@aero.kyushu-u.ac.jp} \\
}

\begin{document}
\maketitle

\begin{abstract}
In mid-2024, asteroid \astname was identified as the first-ever Saturn Trojan through ground-based archival observations and numerical simulations. Trojans, including those associated with Jupiter and other planets, raise important questions about the formation processes of our solar system --- \textit{Were the celestial bodies formed as primordial entities in their current locations, or did they originate from elsewhere?} Additionally, observers have suggested that \astname could be the first active Trojan in our solar system. Exploring this Trojan object with spacecraft may provide direct answers and definitive evidence regarding these questions. However, due to \astname's significant orbital inclination relative to the ecliptic plane and its long orbital period of over 30~years, opportunities to visit this object may be limited to approximately every 15~years. This paper thoroughly investigates potential mission scenarios to the first Saturn Trojan, \astname, to determine the necessary launch window and spacecraft specifications. First, by assuming a ballistic flight using chemical engines, optimal sequences of events, including (powered) gravity-assist maneuvers and deep-space maneuvers, are identified through a meta-heuristic global trajectory optimization algorithm. The analysis indicates that flyby exploration is feasible with a launch window around 2034 and a $\dv$ ranging from 92~m/s to 1041~m/s within an 11-year mission duration, while a rendezvous can be achieved with a departure around 2035 and a $\dv$ of 2--3~km/s. Specifically, the itinerary via Saturn requires a $\dv$ of 2~km/s and a flight time of 24.6~years, while the route via Jupiter results in a $\dv$ of 3~km/s and a flight time of 13.4 years. Given that the orbital motion of outer solar system objects is relatively slow, low-thrust propulsion, which gradually accelerates the spacecraft, proves to be effective. Consequently, low-thrust trajectories to \astname were examined. The results demonstrate that a rendezvous can be accomplished with nearly the same departure epoch, time of flight, and $\dv$ as in the ballistic flight, suggesting that low-thrust propulsion is highly compatible with the rendezvous scenario, as it significantly reduces the propellant mass fraction.
\end{abstract}


\section{Introduction}\label{sec1}
The Lagrange points are five equilibrium positions in celestial mechanics where the gravitational forces of two massive bodies and the centrifugal force are balanced. When the mass ratio of the two primary bodies is sufficiently large, the $L_4$ (leading) and $L_5$ (trailing) points are known to be stable, indicating that celestial objects can be captured in these regions. Small celestial bodies that remain near the $L_4$ or $L_5$ points are referred to as Trojans. To date, numerous Trojans have been discovered, most of which belong to the Sun--Jupiter system.

Deepening our understanding of Trojans plays a prominent role in elucidating the formation processes of our solar system and the origins of life. Recent hypotheses propose that Earth and other terrestrial planets formed from rocky and icy planetesimals were scattered during the migration of the giant planets \cite{bib:nice1, bib:nice2, bib:tack}. The Trojans of the giant planets, such as Jupiter and Neptune, are believed to have been trapped in their current orbits during this planetary migration \cite{bib:nice2}. Earth Trojans, which were discovered only recently in 2011 and 2022, may have originated from planetesimals that contributed to the formation of Earth and the Moon \cite{bib:et}.

Despite that Saturn is the second largest planet in the solar system, no Trojan asteroids belonging to the Sun--Saturn system had been recognized until recently. In November~2024, Hui et al.~\cite{bib:uo14} reported that the asteroid \astname, which was discovered on October 23, 2019, is the first known Trojan of Saturn. Ground-based archival observations and $N$-body numerical simulations have confirmed that \astname has remained in the vicinity of the $L_4$ point for 2000~years and is expected to remain in the Trojan state for the next 1000~years. Additionally, the study demonstrated that \astname has been and will continue to be a horseshoe co-orbital of Saturn during other periods (i.e., more than 2000~years ago and 1000~years from now or later). Consequently, \astname is classified as a ``transient'' Saturn Trojan because it remains in a Trojan state for a relatively short duration compared to other Trojans. Hui et al.~\cite{bib:uo14} measured the properties of \astname and found that it resembles blue Kuiper Belt objects, similar to the Trojans of Jupiter and Neptune. Furthermore, the findings suggest that \astname could be the first active Trojan in the solar system, although no evidence has been found to support this claim yet.

It is anticipated that many mysteries surrounding the Trojans, which remain elusive when relying solely on ground-based observations, can be unraveled through on-site exploration using spacecraft. NASA's Lucy mission, launched on October 16, 2021, is scheduled to conduct consecutive flybys of multiple Jupiter Trojans located at both the $L_4$ and $L_5$ points \cite{bib:lucy}. The Japan Aerospace Exploration Agency (JAXA) had proposed a landing-and-sampling mission to the Jupiter Trojans as part of its Oversize Kite-craft for Exploration and AstroNautics in the Outer Solar system (OKEANOS) mission \cite{bib:okeanos1, bib:okeanos2}, although this proposal was ultimately canceled due to budgetary constraints. Three options were considered for this mission: single rendezvous, multiple rendezvous, and sample return, all aimed at elucidating the material characteristics of the Jupiter Trojans. High-precision orbit determination of the Trojans will also be achievable after arrival, using radio communication between the spacecraft and Earth. Lei et al.~\cite{bib:et_exp} have proposed a rendezvous mission scenario for the first Earth Trojan, 2010~TK$_7$, although no specific mission has been launched yet. Such field investigations could reveal the origins of the Trojans, potentially providing definitive physical evidence to resolve the ongoing debate regarding solar system formation processes, which are currently based primarily on hypotheses and simulations. Furthermore, in the case of \astname, traces of cometary activity can be directly investigated on-site.

To reach the Trojans of the outer planets, advanced skills in trajectory design and engineering technologies are essential, as these asteroids orbit the Sun with significant orbital radii and inclinations. In the case of Jupiter Trojan explorations, the Lucy spacecraft conducts multiple flybys of the Jupiter Trojans in a heliocentric orbit, utilizing gravity assists from Earth \cite{bib:lucy2}. These flybys must be timed to coincident with the moments when the asteroids cross the ecliptic plane, specifically at the ascending or descending nodes (hereafter referred to simply as ``nodes''). In the case of rendezvous, which was planned in the OKEANOS mission, a gravity assist from Jupiter is necessary in order to align the spacecraft's orbital plane with that of the target Trojan. This gravity assist must occur near the nodes of the target \cite{bib:jga1, bib:jga2}. Since a rendezvous mission to the Trojans of Jupiter and beyond requires a substantial amount of $\dv$, the use of electric propulsion is advantageous. However, extremely large solar panels are needed to capture sufficient solar power in the outer solar system. To address this challenge, Japan has proposed a cutting-edge technology known as the solar power sail. In this innovative concept, multiple thin-film flexible solar arrays are attached to a solar sail, enabling the generation of solar power within a lightweight system \cite{bib:okeanos1, bib:jga2}.

Designing a Saturn Trojan exploration mission involves similar challenges. Because \astname has a large inclination of over 30~degrees, flyby opportunities will be restricted to the epochs when it crosses its nodes. A gravity assist from Saturn could be essential in a rendezvous scenario in order to transition to the Trojan's orbital plane. Since both \astname and Saturn have an orbital period of approximately 30~years, it is anticipated that launch opportunities will be significantly limited, likely occurring every 15~years. Furthermore, reaching the Saturn region is a complex challenge. As demonstrated by the Cassini mission, gravity assists from inner planets and deep-space maneuvers (DSMs) are effective in reducing the $\dv$ requirements \cite{bib:cassini}. Numerous combinations of intermediate flybys must be taken into account to design such kind of trajectories. Consequently, the possible scenarios for the exploration of \astname and the associated $\dv$ budget remain uncertain.

In this paper, a comprehensive mission analysis for the exploration of the first-ever Saturn Trojan, \astname, is presented. Optimal mission configurations, including the intermediate flyby sequence, launch window, $\dv$, and time of flight (TOF), are examined for both flyby and rendezvous scenarios. Initially, assuming a ballistic flight utilizing chemical engines, global trajectory optimization is conducted using a meta-heuristic algorithm. The results indicate that the earliest launch opportunity will occur soon (in the early 3030s), with the next opportunity arising 10~years later. In the case of rendezvous, a significant $\dv$ is required in the outer solar system region. Due to the slow orbital motion of celestial objects in this area, low-thrust propulsion, which gradually accelerates the spacecraft with fewer propellant consumption, proves to be highly effective. Consequently, low-thrust trajectories to \astname are also explored by refining the ballistic trajectories. Since there is currently no actual mission plan to \astname, a scaled low-thrust trajectory model is developed in which the spacecraft mass and thrust are normalized. This model allows for a comprehensive understanding of the requirements for the low-thrust mission without the need to assume specific launch mass and thrust magnitude. The findings will be instrumental in determining the scale of the spacecraft when an actual mission plan is formulated in the near future.

The remainder of this paper is organized as follows. First, Section~\ref{sec2} analyzes the orbital motion of \astname over the next century to provide dynamical insights for mission design. Section~\ref{sec3} begins with the formulation of the ballistic trajectory optimization model. Utilizing a meta-heuristic global optimization algorithm, optimal mission scenarios are computed for both the flyby and rendezvous cases. In Sec.~\ref{sec4}, a scaled low-thrust trajectory model is developed, followed by a parametric study on normalized mass and thrust. Finally, Sec.~\ref{sec5} concludes the paper.

\section{Orbit Analysis}\label{sec2}
In the original paper on the discovery of \astname, authored by Hui et al.~\cite{bib:uo14}, the long-term evolution of its orbit over thousands of years is evaluated. However, from an exploratory perspective, the orbital motion over the next few decades is of greater significance, which is not addressed in \cite{bib:uo14}. Although \astname is currently situated at the Sun--Saturn $L_4$ point, its orbit around this point is subject to fluctuations. Therefore, in this section, the short-term orbital motion of \astname over the next century is analyzed to gain dynamic insights that will be applied in the subsequent mission design sections.

\begin{table}[!b]
	\centering
	\caption{Orbital elements of \astname in the heliocentric J2000 ecliptic reference system \cite{bib:uo14}}
	\label{tab:oe}
	\begin{tabular}{lcc}
		\hline
		Orbital element  &  Symbol  &  Value \\
		\hline
		Semimajor axis [au]  &  $a$  &  9.7956923(71) \\
		Eccentricity & $e$ & 0.23639028(48) \\
		Inclination [deg]  &  $i$  &  32.8291888(27) \\
		Longitude of perihelion [deg]  &  $\varpi$  &  28.808276(30) \\
		Argument of perihelion [deg]  &  $\omega$  &  144.167993(30)  \\
		Longitude of ascending node [deg]  &  $\Omega$  &  244.6402830(39)  \\
		Mean anomaly [deg]  &  $M_0$  &  52.553978(68)  \\
		Reference epoch [TDT]  &  $t_0$  &  2024-Apr-04 \\
		\hline
	\end{tabular}
\end{table}

\begin{figure}[!b]
	\centering
	\includegraphics[width=0.95\linewidth]{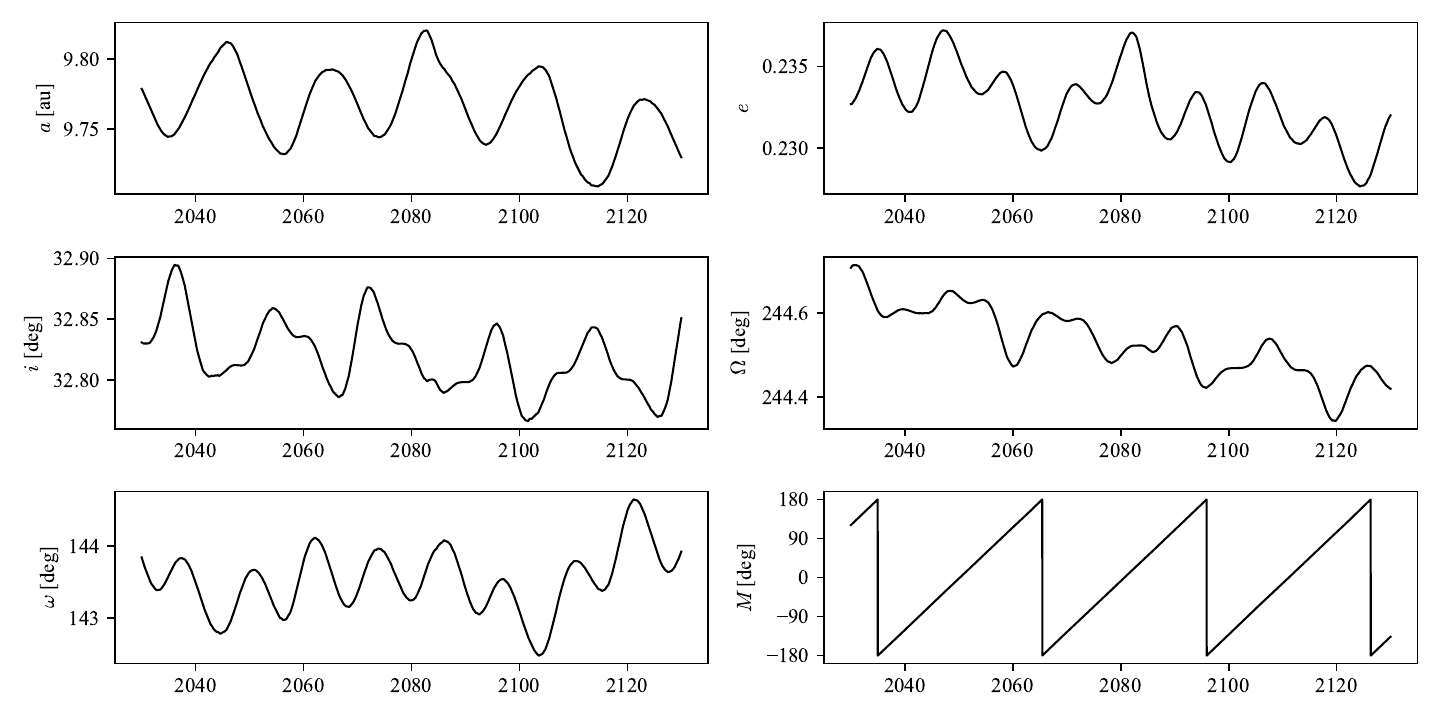}
	\caption{Osculating orbital elements of \astname over the next century.}
	\label{fig:oe}
\end{figure}

Table~\ref{tab:oe} shows the best-fit orbital elements of \astname, as measured by Hui et al.~\cite{bib:uo14}. These values indicate that the orbit of \astname is closer to a circular shape  than to an elliptical one. Similar to the Jupiter Trojans, \astname exhibits a significant orbital inclination relative to the ecliptic plane, which makes it difficult for a spacecraft to reach the object. Figure~\ref{fig:oe} shows osculating (i.e., variation of) orbital elements of \astname from 2030 to 2130 calculated using the SPICE kernel provided by the JPL Horizons System. Although some orbital elements undergo short-period oscillations of approximately 20 years, their amplitude is minimal, indicating that the orbit of \astname remains nearly constant over the next century.

\begin{figure}[!b]
	\centering
	\begin{minipage}{0.49\linewidth}
		\includegraphics[width=\linewidth]{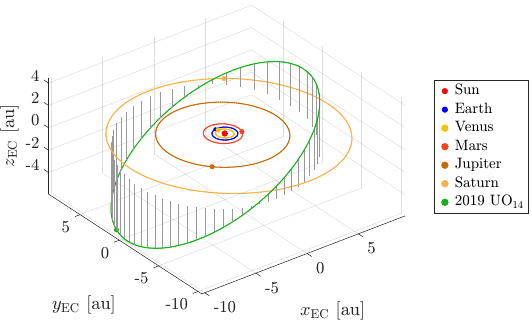}
		\subcaption{J2000EC as of January 1, 2030}
		\label{fig:orbit_ec}
	\end{minipage}
	\begin{minipage}{0.49\linewidth}
		\includegraphics[width=\linewidth]{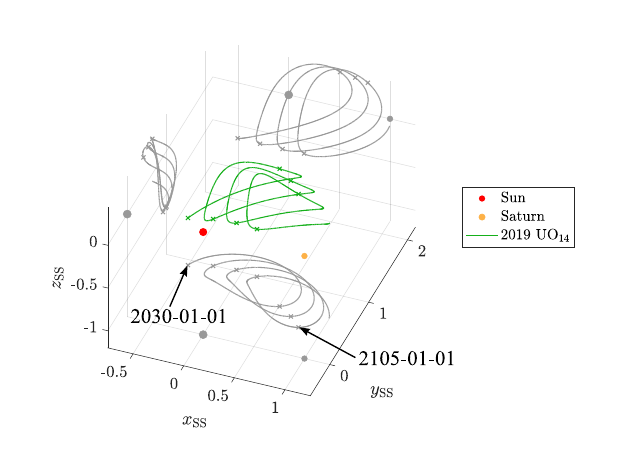}
		\subcaption{Sun--Saturn frame}
		\label{fig:orbit_ss}
	\end{minipage}
	\caption{Orbit plots of \astname}
	\label{fig:orbit}
\end{figure}

Figure~\ref{fig:orbit} illustrates the orbit of \astname plotted in the ecliptic J2000 (J2000EC) coordinate system and in the Sun--Saturn-fixed (SS) rotating frame. Note that the coordinates in the SS frame are normalized by the Sun--Saturn distance. The cross markers in Fig.~\ref{fig:orbit_ss} are plotted at 15-year intervals, ranging from January 2030 to January 2120. This figure confirms that \astname remains in proximity to the Sun--Saturn L4 point and gradually approaches Saturn over time. Due to \astname's highly inclined orbit, gravity-assist maneuvers are crucial for executing a rendezvous mission. Utilizing a Saturn flyby to increase inclination is more advantageous with a later launch, as the journey from Saturn to \astname becomes shorter. In other words, launching our spacecraft at the earliest opportunity requires significantly more effort (e.g., $\dv$ and TOF) to achieve a rendezvous with \astname.

\begin{figure}[!b]
	\centering
	\includegraphics[width=0.5\linewidth]{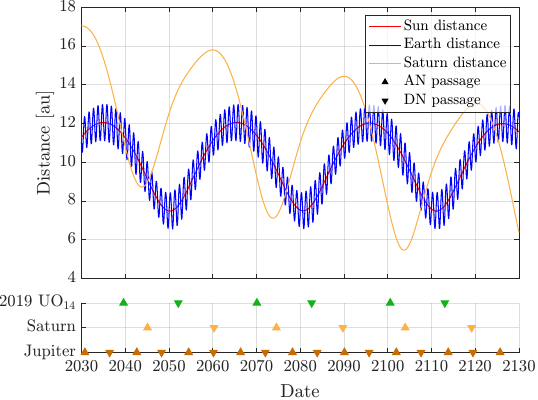}
	\caption{Geometric characteristics of \astname.}
	\label{fig:geom}
\end{figure}

Figure~\ref{fig:geom} shows geometric characteristics of \astname. Due to the short-term stability of \astname's orbit, both the solar distance and the Earth distance exhibits stable oscillations in response to the orbital motions of \astname and Earth. The maximum solar and Earth distances are 12~au and 13~au respectively, indicating significant requirements for the design of electrical power control, thermal control, and communication systems. This figure also confirms that \astname approaches Saturn over time. Triangular and inverted triangular markers represent, respectively, the epochs when each celestial body crosses the ascending node (AN) and descending node (DN) of \astname. The earliest epoch when \astname intersects the ecliptic plane, which is a critical timing for a flyby mission, occurs on August 20, 2039. Given that it takes approximately 7 to 11 years to reach the Saturn region, as demonstrated in previous missions and studies \cite{bib:cassini, bib:mga1dsm}, it is imperative to initiate the flyby mission promptly if this opportunity is to be pursued. Saturn will cross the AN of \astname six years later, on April 12, 2045, which is advantageous for aligning the spacecraft's orbital plane with that of \astname through a Saturn gravity assist (SGA). Since both \astname and Saturn have an orbital period of about 30~years, the opportunities for a \astname flyby within the ecliptic plane and for inclination raising via an SGA occur approximately every 15~years. Additionally, because the analyses in Sec.~\ref{sec3} have shown that a Jupiter gravity assist (JGA) is also effective for inclination raising, the epochs when Jupiter crosses the AN and DN of \astname are plotted in Fig.~\ref{fig:geom}. Considering that Jupiter has an orbital period of 12~years, JGA opportunities arise every 6~years. However, it is important to note that not all of JGA opportunities are beneficial, as the characteristics of the Jupiter--\astname transfer trajectory vary depending on the relative position of Jupiter and \astname. In the next section, the transfer trajectory to \astname will be designed by leveraging the insights into the orbital dynamics of Jupiter, Saturn, and \astname.

\section{Ballistic Trajectories}\label{sec3}

\subsection{Trajectory Model}\label{sec31}
To date, design methods for interplanetary trajectories have been extensively studied by numerous researchers \cite{bib:mga1dsm, bib:trajopt1, bib:trajopt2, bib:trajopt3}. One of the oldest yet still effective methods is the multiple gravity-assist (MGA) transcription, which incorporates gravity-assist maneuvers at multiple intermediate bodies. Since the trajectory between two sequential bodies is fully ballistic, it can be solved using Lambert's problem. $\dv$ maneuvers are permitted only at the periapsis of each intermediate body. The flybys must be designed so that the periapsis altitude does not fall below the allowable threshold. The MGA problem can be formulated as an unconstrained single-objective optimization problem with a limited number of decision variables, for which heuristic algorithms such as genetic algorithms, differential evolution, and particle swarm optimization are applicable. In practice, it is restrictive to prohibit orbital maneuvers during interplanetary cruise. The multiple gravity-assist with one deep-space maneuver (MGA-1DSM) model allows for a single rocket burn in the coast arc between each pair of celestial bodies, enhancing the flexibility of mission design without significantly increasing computational costs. Previous studies on MGA-1DSM problems have typically focused solely on DSMs and have not incorporated powered flybys (P-FBs). However, it has been turned out that permitting both DSMs and P-FBs tends to yield superior solutions. Therefore, assuming a patched-conic approximation, an extended model called multiple powered gravity-assist with one deep-space maneuver (MPGA-1DSM) is developed as follows.

\begin{figure}[!b]
	\centering
	\includegraphics[width=0.5\linewidth]{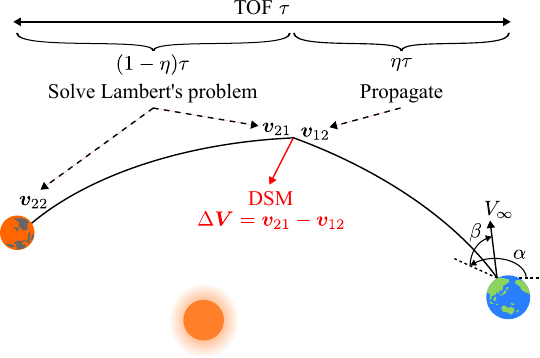}
	\caption{MGA-1DSM trajectory design model.}
	\label{fig:mga_1dsm}
\end{figure}

Figure~\ref{fig:mga_1dsm} illustrates the trajectory design model. For each trajectory leg between a pair of celestial bodies, including the starting body (i.e., Earth), intermediate bodies, and the arrival body (\astname in the present study), the following decision variables are defined:
\begin{align}
	\bm{Y}_i = [V_\infty, \alpha, \beta, \tau, \eta]_i
\end{align}
where $V_\infty$, $\alpha$, and $\beta$ represent, respectively, the magnitude, azimuth, and elevation angle of the outgoing $V_\infty$ vector at the $i$th body; $\tau$ is the TOF between the $i$th and $(i+1)$th bodies; $\tau \in [0, 1]$ indicates the timing of DSM; and $i = 1, \cdots, N$ is the trajectory leg index. With the departure epoch denoted as $t_0$, the complete decision vector for the MPGA-1DSM problem is defined as
\begin{align}
	\bm{X} = [t_0, \bm{Y}_1, \cdots, \bm{Y}_N]
\end{align}
where $N$ is the number of trajectory legs. The epochs when the spacecraft reaches the $j$th body is given by
\begin{align}
	\begin{aligned}
		t_1 &= t_0 &\quad (j = 1)\\
		t_j &= t_0 + \sum_{i=1}^{j  - 1} \tau_i &\quad (j \ge 2)
	\end{aligned}
\end{align}
for $j = 1, \cdots, N + 1$. The position and the velocity of the $j$th body, $\bm{r}_{p, j}$ and $\bm{v}_{p, j}$, can be computed using the SPICE Toolkit, corresponding ephemeris files, and the epochs $t_j$. The outgoing velocity vector from the $i$th body is
\begin{align}
	\bm{v}_{11, i} = \bm{v}_{p, i} + \bm{V}_{\infty \text{-out}, i}
\end{align}
where
\begin{align}
	\bm{V}_{\infty \text{-out}, i} = v_{\infty, i} \left[
		\begin{matrix}
			\cos\alpha_i \cos\beta_i \\
			\sin\alpha_i \cos\beta_i \\
			\sin\beta_i
		\end{matrix}
	\right]
\end{align}
Starting from the $i$th body with the velocity $\bm{v}_{11, i}$, the spacecraft's trajectory is propagated for $\eta \tau$ under the assumption of a heliocentric two-body problem. This yields the position vector $\bm{r}_{12, i}$ and the velocity vector $\bm{v}_{12, i}$ at the point where the DSM occurs. Next, Lambert's problem is solved between the DSM position $\bm{r}_{12, i} = \bm{r}_{21, i}$ and the $(i+1)$th body with a transfer time $(1 - \eta) \tau$, which gives the starting velocity $\bm{v}_{21, i}$ at the DSM position and the arrival velocity $\bm{v}_{22, i}$ at the $(i+1)$th body. The difference in velocity vectors at the DSM position corresponds to the DSM $\dv$.
\begin{align}
	\dv_{\mathrm{DSM}, i} = \| \bm{v}_{21, i} - \bm{v}_{21, i} \|
\end{align}
The incoming hyperbolic velocity at the $(i+1)$th body is
\begin{align}
	\bm{V}_{\infty\text{-in}, i + 1} = \bm{v}_{22, i} - \bm{v}_{p, i + 1}
\end{align}
For $i = 1, \cdots, N-1$, the incoming and outgoing hyperbolic velocities, $\bm{V}_{\infty\text{-in}, i + 1}$ and $\bm{V}_{\infty\text{-out}, i + 1}$, must satisfy a constraint so that the flyby is feasible.

\begin{figure}
	\centering
	\includegraphics[width=\linewidth]{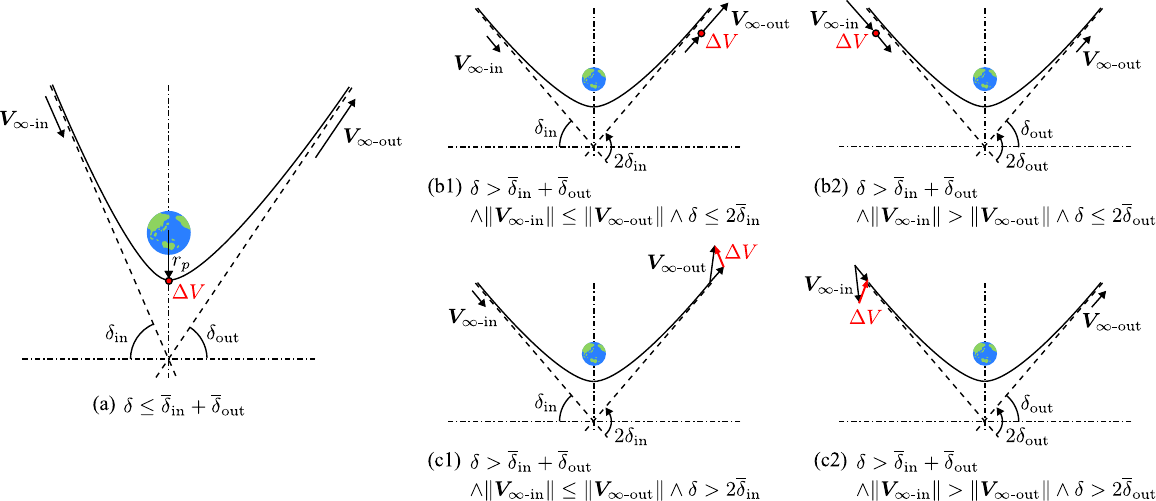}
	\caption{Powered flyby model.}
	\label{fig:p_fb}
\end{figure}

Given that the spacecraft passes through the periapsis at a distance of $r_{p, i + 1}$, the deflection angle for each infinity velocity is expressed as follows:
\begin{align}
	\begin{aligned}
		\delta_{\mathrm{in}, i + 1} &= \sin^{-1}(1 / e_{\mathrm{in}, i + 1}) \\
		\delta_{\mathrm{out}, i + 1} &= \sin^{-1}(1 / e_{\mathrm{out}, i + 1}) \\
	\end{aligned}
	\label{eq:deflec}
\end{align}
with
\begin{align}
	\begin{aligned}
		e_{\mathrm{in}, i + 1} &= 1 + \frac{r_{p, i + 1}}{\mu_{i + 1}} \| \bm{V}_{\infty \text{-in}, i + 1} \|^2 \\
		e_{\mathrm{out}, i + 1} &= 1 + \frac{r_{p, i + 1}}{\mu_{i + 1}} \| \bm{V}_{\infty \text{-out}, i + 1} \|^2
	\end{aligned}
	\label{eq:ecc}
\end{align}
where $\mu_{i +1}$ represents the gravitational parameter of the $(i + 1)$th body. Typically, a minimum periapsis distance is defined to ensure that the flyby can be completed safely. In previous studies (e.g., \cite{bib:mga1dsm}), a constraint was imposed such that the periapsis distance does not fall below a specified lower limit. To formulate an unconstrained optimization problem, a penalty function was employed, which increases the objective value in response to any violation of the periapsis distance. In the present study, a $\dv$ is permitted at each intermediate flyby to let the flyby be feasible; thus, the periapsis distance constraint is implicitly incorporated by minimizing the total $\dv$. Depending on the configuration of the $V_\infty$ vector diagram, the P-FB $\dv$ is calculated in three scenarios as illustrated in Fig.~\ref{fig:p_fb}. For the minimum periapsis distance $\overline{r}_{p, i + 1}$, the maximum deflection angle $\overline{\delta}_{i + 1} = \overline{\delta}_{\mathrm{in}, i + 1} + \overline{\delta}_{\mathrm{out}, i + 1}$ can be determined using Eqs.~(\ref{eq:deflec}) and (\ref{eq:ecc}). If the actual deflection angle
\begin{align}
	\delta_{i + 1} = \cos^{-1} \left(
		\frac{\bm{V}_{\infty \text{-in}, i + 1}}{\| \bm{V}_{\infty \text{-in}, i + 1} \|} \cdot \frac{\bm{V}_{\infty \text{-out}, i + 1}}{\| \bm{V}_{\infty \text{-out}, i + 1} \|}
	\right)
	\label{eq:deflec2}
\end{align}
is smaller than its upper limit $\overline{\delta}_{i + 1}$, the Oberth effect can be utilized by conducting the $\dv$ maneuver at the periapsis passage. The P-FB $\dv$ in this case is given by
\begin{align}
	\dv_{\mathrm{P-FB}, i + 1} = \left|
		\sqrt{\| \bm{V}_{\infty \text{-in}, i + 1} \|^2 + \frac{2 \mu_{i + 1}}{r_{p, i + 1}}} - \sqrt{\| \bm{V}_{\infty \text{-out}, i + 1} \|^2 + \frac{2 \mu_{i + 1}}{r_{p, i + 1}}}
	\right|
	\label{eq:oberth}
\end{align}
The actual periapsis distance $r_{p, i + 1}$ in Eq.~(\ref{eq:oberth}) can be computed by numerically solving the following equation: $\delta_{i + 1} = \delta_{\mathrm{in}, i + 1} + \delta_{\mathrm{out}, i + 1}$. If the actual deflection angle, as expressed in Eq.~(\ref{eq:deflec2}), exceeds the upper limit $\overline{\delta}_{i + 1}$, the Oberth effect cannot be utilized. In this case, the spacecraft executes a standard flyby (i.e., without a $\dv$) to alter the infinity velocity vector. A $\dv$ is applied either before or after the flyby to compensate for the $V_\infty$ mismatch. If the actual deflection angle $\delta_{i + 1}$ is less than $2 \max (\overline{\delta}_{\mathrm{in}, i + 1}, \overline{\delta}_{\mathrm{out}, i + 1})$, it is sufficient to accelerate after the flyby or decelerate before the flyby in the direction of $V_\infty$, as illustrated in cases (b1) and (b2). The required $\dv$ is
\begin{align}
	\dv_{\mathrm{P-FB}, i + 1} = \left|
		\| \bm{V}_{\infty \text{-out}} \| - \| \bm{V}_{\infty \text{-in}} \|
	\right|
	\label{eq:p_fb1}
\end{align}
If $\delta_{i + 1}$ is greater than $2 \max (\overline{\delta}_{\mathrm{in}, i + 1}, \overline{\delta}_{\mathrm{out}, i + 1})$, the $\dv$ must be applied to further deflect the $V_\infty$ vector as indicated in cases (c1) and (c2). This maneuver can be accomplished by
\begin{align}
    \begin{aligned}
	   &\dv_{\mathrm{P-FB}, i + 1} \\
       &= \sqrt{\| \bm{V}_{\infty \text{-in}, i + 1} \|^2 + \| \bm{V}_{\infty \text{-out}, i + 1} \|^2 - 2 \| \bm{V}_{\infty \text{-in}, i + 1} \| \cdot \| \bm{V}_{\infty \text{-out}, i + 1} \| \cos\left\{\delta_{i + 1} - 2 \max(\overline{\delta}_{\mathrm{in}, i + 1}, \overline{\delta}_{\mathrm{out}, i + 1})\right\}}
    \end{aligned}
	\label{eq:p_fb2}
\end{align}
In the case of rendezvous missions, the following $\dv$ is needed at the arrival:
\begin{align}
	\dv_\mathrm{arr} = \left\{
		\begin{matrix}
			\| \bm{V}_{\infty \text{-in}, N + 1} \| & \quad \textrm{for a rendezvous mission} \\
			0 & \quad \textrm{for a flyby mission}
		\end{matrix}
	\right.
\end{align}

Thus, the objective function can be written as the sum of the $\dv$s as follows:
\begin{align}
	f(\bm{X}) = \sum_{i = 1}^{N} \dv_{\mathrm{DSM}, i} + \sum_{i = 1}^{N - 1} \dv_{\mathrm{P-FB}, i + 1} + \dv_\mathrm{arr} 
\end{align}
In practice, it is beneficial to incorporate an upper bound on the TOF using a penalty method as follows:
\begin{align}
	g(\bm{X}) = \left\{
		\begin{matrix}
			0 & \textrm{if} \sum_{i = 1}^{N}\tau_i \le \tau_\mathrm{max} \\
			w \left(
				\sum_{i = 1}^{N}\tau_i - \tau_\mathrm{max}
			\right) & \textrm{if} \sum_{i = 1}^{N}\tau_i > \tau_\mathrm{max} 
		\end{matrix}
	\right.
\end{align}
where $w$ is the weight factor and $\tau_\mathrm{max}$ is the allowable total flight time. Additional constraints can be addressed by incorporating them into the penalty function. For instance, previous studies \cite{bib:mga1dsm, bib:trajopt3} have included a penalty for low-velocity flybys, where the spacecraft's incoming velocity is insufficient, leading to the collapse of the hyperbolic trajectory assumption. However, in the present study, this issue can be avoided by establishing a sufficiently high lower bound on $V_\infty$. While the penalty method can be beneficial, excessive penalties may lead to an unstable behavior, as the objective value becomes entangled with numerous penalties in a single scalar. Therefore, this study focuses solely on the maximum TOF constraint. Consequently, the trajectory design problem is formulated as an unconstrained (box-bounded) single-objective optimization problem as follows:
\begin{align}
		\min_{\bm{X}} f(\bm{X}) + g(\bm{X}) \quad \textrm{subject to} \quad  \bm{X}_\mathrm{lb} \le \bm{X} \le \bm{X}_\mathrm{ub}
		\label{eq:trajopt}
\end{align}
where $\bm{X}_\mathrm{lb}$ and $\bm{X}_\mathrm{ub}$ are the lower bound and the upper bound on the decision vector.

\subsection{Global Optimization Method}\label{sec32}
Because Eq.~(\ref{eq:trajopt}) is formulated as a box-bounded single-objective optimization model, population-based global optimization methods such as genetic algorithms, differential evolution, and particle swarm optimization (PSO) are applicable. In these methods, a population consisting of random decision vectors is initially generated, and then the population is evolved over several generations using various algorithms to identify a better solution. Population-based methods can explore the solution space globally without requiring an initial guess as well as gradient information. However, a substantial number of individuals are needed in the population to thoroughly investigate the solution space, which necessitates numerous function evaluations and consequently increases computational costs. Furthermore, the optimization results can vary with each run due to the stochastic nature of the process, meaning that the global optimality is not guaranteed. One effective solution to this issue is search space pruning. Vasile and De Pascale~\cite{bib:trajopt1} proposed evolutionary branching, in which the optimization problem is addressed through an evolutionary algorithm, and the solution space is partitioned into smaller subdomains based on the optimization results. The optimization is then repeated within the updated (pruned) solution space until no further improvements in fitness values are observed. Although this method successfully identifies optimal interplanetary trajectories for various scenarios, it is important to note that it may overlook regions where better solutions exist during the pruning process. Englander et al.~\cite{bib:mga1dsm} proposed a different approach that hybridizes trajectory optimization in the inner loop with the flyby sequence optimization in the outer loop. In this method, a fixed value was employed for the search space depending on the combination of flyby bodies.

\begin{figure}[!b]
	\centering
	\includegraphics[width=\linewidth]{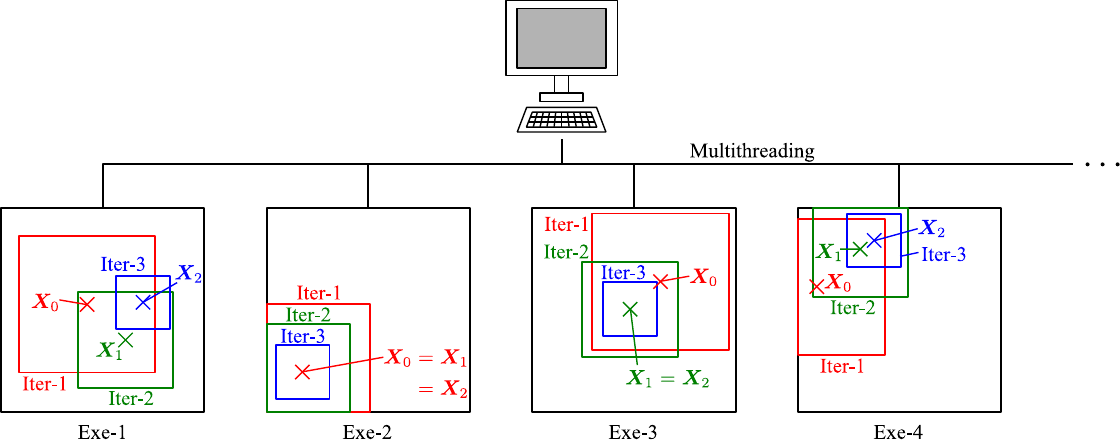}
	\caption{Global optimization through parallelized search space pruning.}
	\label{fig:gopt}
\end{figure}

Considering that computer performance has improved significantly over the past decade, a more comprehensive search is now possible. In the present study, a meta-heuristic global optimization algorithm that leverages modern computer capabilities is presented. For a given sequence of planetary flybys, global optimization is conducted using the method sketched in Fig.~\ref{fig:gopt}. First, the optimization problem defined by Eq.~(\ref{eq:trajopt}) is solved with user-specified initial bounds. In this study, these initial bounds are determined based on the sequence of intermediate bodies, as detailed in the Appendix. The solution obtained is denoted as $\bm{X}_0$. Next, the solution domain is reduced by a scale factor $\lambda \in [0, 1]$ around $\bm{X}_0$. Within this new subdomain, Eq.~(\ref{eq:trajopt}) is solved again (iter-1), yielding a new solution $\bm{X}_1$. The solution domain is further contracted by the scale factor $\lambda$ around the better solution, either $\bm{X}_0$ or $\bm{X}_1$, followed by solving Eq.~(\ref{eq:trajopt}) again (iter-2), which produces the next solution $\bm{X}_2$. This iterative process continues until the user-specified maximum number of iterations is reached. This calculations, involving iterative pruning, are parallelized in a multi-threaded environment (exe-1, exe-2, ...). Because the evolutionary algorithm described in the following paragraph incorporates stochastic processes, the initial solution ($\bm{X}_0$) and subsequent solutions ($\bm{X}_1, \bm{X}_2, \cdots$) converge to different values with each execution (i.e., in each thread). Consequently, different optima can be efficiently obtained in parallel. The probability of identifying the global optimum increases by ensuring a sufficient number of iterations and executions.

For each optimization of Eq.~(\ref{eq:trajopt}), a combined PSO--MBH algorithm is employed. Initially, Eq.~(\ref{eq:trajopt}) is solved using conventional PSO \cite{bib:pso1, bib:pso2}. In PSO, a number of particles are randomly generated in the solution space. These particles move within the solution space at a certain velocity over several generations. All particles share the global best, which is the best (i.e., minimum-objective) solution found across all particles and the generations. Additionally, each particle retains its local best, which is the best solution that it has experienced. The velocity vector of each particle is updated toward an intermediate direction between the global best and the local best, which is determined by random weight factors. PSO facilitates an efficient and global search of the solution space without requiring an initial guess. However, since PSO does not utilize gradient information, the updates toward the potentially best solution may lack precision. Consequently, the solution obtained through PSO may not be truly optimal. To address this issue, the PSO-derived solution is refined using a gradient-based local optimization method. In this study, the Sequential Quadratic Programming (SQP) method is employed. This solution is further refined through Monotonic Basin Hopping (MBH). In MBH, the initial guess is generated by randomly perturbing the previous solution, and it is optimized using a gradient-based method, which is repeated until no further improvement is observed \cite{bib:mbh}.

The summary of the meta-heuristic algorithm is presented as follows. The optimization problem defined by Eq.~(\ref{eq:trajopt}) is solved using a hybrid global--local optimization algorithm that combines PSO, SQP, and MBH. This subproblem is iteratively solved while progressively narrowing the solution space around the best solution, which is managed independently in a parallelized environment. Analyses in the subsequent subsections confirm that, for an interplanetary trajectory involving five intermediate flybys, a single execution comprising five iterations (i.e., solving for $\bm{X}_0, \cdots, \bm{X}_4$) can be completed in about 10~minutes using an iMac (Apple M1, MacOS~14.6.1) and a Dell Inspiron 15 (AMD Ryzen 7 5700U, Ubuntu 20.04). Note that the optimization program was written in C++ and compiled to enhance computational speed. This outcome suggests that, for instance, if 30~threads are available, 30 independent executions, each involving five iterations, can be completed within 10 minutes. As detailed in the following subsections, the number of sequences including a maximum of five intermediate flybys does not exceed 200. Therefore, global optimization incorporating all potential flyby sequences with $30 \times 5$ trials for each can be completed within 1 or 2 days, without the need for outer-loop optimization as presented in \cite{bib:mga1dsm}. The proposed method is useful in that the parallel computation described in Fig.~\ref{fig:gopt} can be executed across multiple hardware platforms, as the executions are independent of one another. In other words, if a common setting file is shared among several computers, many-threaded computing becomes possible. In fact, the author used three computers with standard specifications (MacBook Pro with 4 CPU cores and 8 threads; iMac with 4 CPU cores and 8 threads; Dell Inspiron 15 with 8 cores and 16 threads) to establish a 32-thread environment. Advanced skills such as computer clustering are not required. Parallel computing operates simply by executing each tasks in the terminal on each computer.

\subsection{Flyby Mission}\label{sec33}

\begin{table}[!b]
	\centering
	\caption{Properties of the celestial bodies}
	\label{tab:bodies}
	\begin{tabular}{lccccc}
		\hline
		Name  & Acronym  &  Integer code  &  Gravitational parameter   &  Radius   &  Minimum flyby altitude  \\
        & & & [m$^3$/s$^2$] & [km] & [km] \\
		\hline
		Null  &  --  &  -1  &  --  & --  &  -- \\
		Earth  &  E  &  0  &  3.986004418$\times$10$^{14}$  &  6378  &  637.8 \\
		Venus  &  V  &  1  &  3.24859$\times$10$^{14}$  &  6052  &  605.2 \\
		Mars  &  M  &  2  &  4.2828$\times$10$^{13}$  &  3397  &  339.7 \\
		Jupiter  &  J  &  3  &  1.26686534$\times$10$^{17}$  &  71492  &  571936  \\
		Saturn  &  S  &  4  &  3.7931187$\times$10$^{16}$  &  60330  & 6033 \\
		\astname  &  A  &  5  &  --  &  --  &  -- \\
		\hline
	\end{tabular}
\end{table}

Interplanetary trajectories to \astname for a flyby mission is investigated here. First, potential flyby sequences are generated to formulate the trajectory optimization problem. In the design of the flyby mission, four intermediate flybys are allowed at maximum, with at least one intermediate flyby deemed necessary. Venus, Mars, Jupiter, and Saturn are considered as intermediate bodies. An integer code is assigned to each body, as shown in Table~\ref{tab:bodies}. Let the sequence be expressed as EP$_1\cdots$P$_4$A, where P$_k$ denotes the $k$th intermediate body. For example, EVEEJA corresponds to the Earth--Venus--Earth--Earth--Jupiter--\astname sequence with P$_1$=V, P$_2$=E, P$_3$=E, and P$_4$=J\@. Possible planets for each interplanetary flyby are listed in Table~\ref{tab:fb_bodies}. The code -1 indicates that no flyby is performed. For example, [P$_1$, P$_2$, P$_3$, P$_4$] = [1, 0, -1, 2] is equivalent to the EVEMA sequence. Direct transfers from launch to outer planets (i.e., Jupiter and Saturn) require substantial launch energy; therefore, they are excluded from consideration for P$_1$. In addition, Saturn is not included in the flyby mission design because Saturn and \astname are in a co-orbital state, making a flyby through Saturn result in unnecessarily prolonged flight times. All possible sequences are generated from the direct product of the integer codes shown in Table~\ref{tab:fb_bodies}. Duplicate sequences (e.g., [1, -1, 0, 0] and [1, 0, -1, 0] yield the same sequence: EVEEA) are consolidated into a single entry. Furthermore, sequences that involve more than three consecutive flybys by the same planet (e.g., EEEEJA and EMMMMA) are eliminated. Since it is useless to fly through an outer planet and then return to inner planets (e.g., EVEJEA), such sequences are also discarded. As a result, 152 unique sequences were generated. The box bound on the departure $V_\infty$ was set to $[1, 5.1]$~km/s. The upper bound corresponds to $C_3 = 26$~km$^2$/s$^2$, which is the minimum requirement to enter a 1:2 Earth-resonant orbit where a $V_\infty$ leveraging maneuver \cite{bib:dvega} is available. The lower bound was established to prevent low-velocity flybys.

\begin{table}
	\centering
	\caption{Possible combinations of intermediate bodies for flyby missions}
	\label{tab:fb_bodies}
	\begin{tabular}{cc}
		\hline
		Body  &  Integer code  \\
		\hline
		P$_1$  &  [0, 1, 2]  \\
		P$_2$  &  [-1, 0, 1, 2, 3]  \\
		P$_3$  &  [-1, 0, 1, 2, 3]  \\
		P$_4$  &  [-1, 0, 1, 2, 3]  \\
		\hline
	\end{tabular}
\end{table}

As discussed in Sec.~\ref{sec2}, the first opportunity for \astname to corss the ecliptic plane occurs on August 20, 2039, at its AN, while the second opportunity is on February 21, 2052, at its DN\@. Because it may take about 7--11~years to reach the Saturn region, the expected launch windows are projected to be around 2030 and 2043, respectively. This paper investigates two cases corresponding to these launch windows. The bounds on $t_0$ were set to [2025-Jan-01, 2035-Jan-01] in the first case and [2037-Jan-01, 2047-Jan-01] in the second case. The optimization parameter settings are provided in Table~\ref{tab:optset}. The open-source software PAGMO \cite{bib:pagmo} was used to solve the PSO--MBH problem. Default values were employed to construct the PSO and MBH algorithms. The upper limit for the total TOF was set to 4000~days, with a weight factor $w = 10$. A total of 30~independent executions, each consisting of 5 iterations, were conducted for each sequence. The contraction factor was set to $\lambda = 0.6$.

Tables~\ref{tab:opt_fb1} and \ref{tab:opt_fb2} show the results of trajectory design. The P-FBs and DSMs were deemed not to occur if the total $\dv$ is less than 10~m/s.  In the first opportunity, most sequences include Jupiter as the final intermediate body. The JGAs contribute to an increase in the inclination, leading to the arrival at \astname outside the ecliptic plane. Therefore, these sequences are not constrained by the ecliptic plane passage of \astname (August 20, 2039) but rather by that of Jupiter (August 28, 2042). The launch window ranges from late 2033 to late 2034, which is realistic from a project perspective. There are two sequences in which no JGA is used and the flyby of \astname occurs within the ecliptic plane. The spacecraft must be launched very early in 2028 or 2029 (i.e., within four years from the time this paper was written); however, the $\dv$ requirement is not small compared to other sequences. As a result, the analysis indicates that launching the spacecraft in 2033 or 2034 and employing a JGA will be the optimal choice for achieving a flyby observation of \astname at the earliest opportunity. If this opportunity is missed, the next chance will arise between 2039 and 2043, as shown in Table~\ref{tab:opt_fb2}. Because \astname is closer to the Sun during the DN passage than during the AN passage (see Fig.~\ref{fig:orbit_ec}), the requirements for $\dv$ and the launch window are less stringent in the second opportunity compared to the first. It is also noteworthy that the number of in-plane flybys and out-of-plane flybys (with a JGA) are nearly equal in the second opportunity.

\begin{table}[!b]
	\centering
	\caption{Optimization parameter settings}
	\label{tab:optset}
	\begin{tabular}{ccc}
		\hline
		Algorithm  &  Parameter  &  Value  \\
		\hline
		PSO  &  Number of individuals  &  1000 \\
		&  Number of generations  &  2000  \\
		\hline
		MBH  &  Perturbation  &  0.2  \\
		&  Maximum iterations  &  100 \\
		\hline
	\end{tabular}
\end{table}

\begin{table}
	\centering
	\caption{Top 10 sequences for flyby missions at the first opportunity}
	\label{tab:opt_fb1}
	\begin{tabular}{clllccc}
		\hline
		No.  &  Sequence  &  Departure  &  Arrival  &  $\dv$ [m/s]  &  P-FB  &  DSM  \\
		\hline
		1  &  EEMEJA  &  2033-Nov-27  &  2044-Nov-17  &  91.82  &  True  &  True  \\
		2  &  EMMEJA  &  2033-Nov-15  &  2044-Nov-20  &  256.65  &  False  &  True  \\
		3  &  EVEEJA  &  2034-Aug-26  &  2045-Aug-02  &  370.74  &  True  &  False  \\
		4  &  EMEEJA  &  2034-Nov-09  &  2045-Oct-22  &  667.59  &  False  &  True  \\
		5  & EEEJA  &  2034-Dec-28  &  2045-Dec-01  &  750.10  &  False  &  True  \\
		6  &  EEEA  &  2028-Dec-10  &  2039-Aug-21  &  765.75  &  False  &  True  \\
		7  &  EEVEJA  &  2034-Jun-18  &  2044-Dec-27  &  882.22  &  False  &  True  \\
		8  &  EVEEA  &  2029-Mar-06  &  2039-Sep-10  &  968.60  &  False  &  True  \\
		9  &  EMVEJA  &  2034-Dec-28  &  2045-Oct-28  &  974.02  &  False  &  True  \\
		10  &  EVVEJA  &  2034-Mar-03  &  2045-Jan-03  &  1040.97  &  True  &  True \\
		\hline
	\end{tabular}
\end{table}

\begin{table}
	\centering
	\caption{Top 10 sequences for flyby missions at the second opportunity}
	\label{tab:opt_fb2}
	\begin{tabular}{clllccc}
		\hline
		No.  &  Sequence  &  Departure  &  Arrival  &  $\dv$ [m/s]  &  P-FB  &  DSM  \\
		\hline
		1  &  EVEEJA  &  2039-Mar-21  &  2049-Dec-14  &  0.10  &  False  &  False  \\
		2  &  EVEEA  &  2042-Mar-30  &  2052-Mar-04  &  39.40  &  False  &  True  \\
		3  &  EMMEEA  &  2041-Jun-02  &  2052-Jan-11  &  314.24  &  False  &  True  \\
		4  &  EEVEEA  &  2041-Aug-10  &  2052-Feb-25  &  382.68  &  True  &  True  \\
		5  &  EVEJA  &  2038-Nov-13  &  2049-Oct-26  &  463.20  &  True  &  True  \\
		6  & EMEEA  &  2042-Feb-20  &  2052-Feb-29  &  477.37  &  True  &  True  \\
		7  &  EEEJA  &  2038-Dec-21  &  2049-Dec-02  &  617.33  &  True  &  True  \\
		8  &  EEJA  &  2040-Mar-05  &  2051-Feb-16  &  619.51  &  False  &  True  \\
		9  &  EVMVEA  &  2042-Mar-08  &  2052-Jan-28  &  666.28  &  False  &  True  \\
		10  &  EEEA  &  2043-Aug-06  &  2052-Feb-20  &  703.19  &  False  &  True  \\
		\hline
	\end{tabular}
\end{table}

\begin{figure}
	\centering
	\begin{minipage}{0.49\linewidth}
		\includegraphics[width=\linewidth]{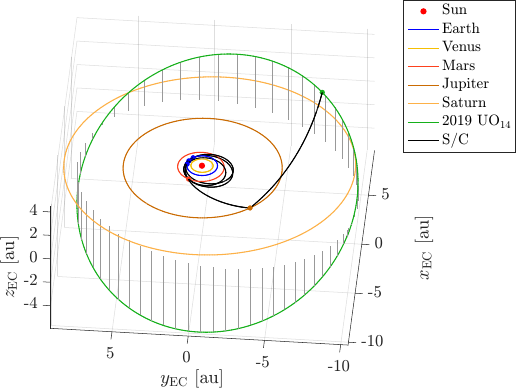}
		\subcaption{Trajectory}
		\label{fig:fb_eemeja_traj}
	\end{minipage}
	\begin{minipage}{0.49\linewidth}
		\includegraphics[width=\linewidth]{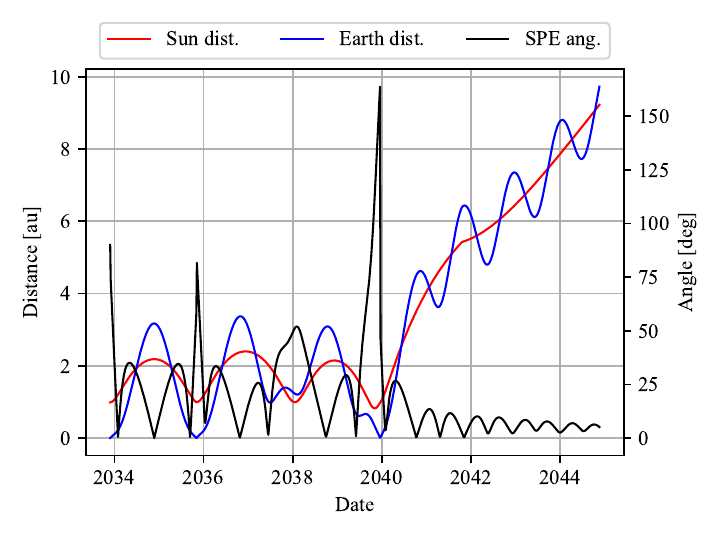}
		\subcaption{Geometry}
		\label{fig:fb_eemeja_geom}
	\end{minipage}
	\caption{Flyby trajectory with EEMEJA sequence (launch on November 27, 2033).}
	\label{fig:fb_eemeja}
\end{figure}

\begin{figure}
	\centering
	\begin{minipage}{0.49\linewidth}
		\includegraphics[width=\linewidth]{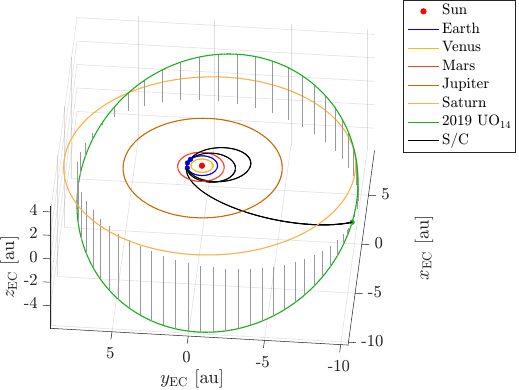}
		\subcaption{Trajectory}
		\label{fig:fb_eeea_traj}
	\end{minipage}
	\begin{minipage}{0.49\linewidth}
		\includegraphics[width=\linewidth]{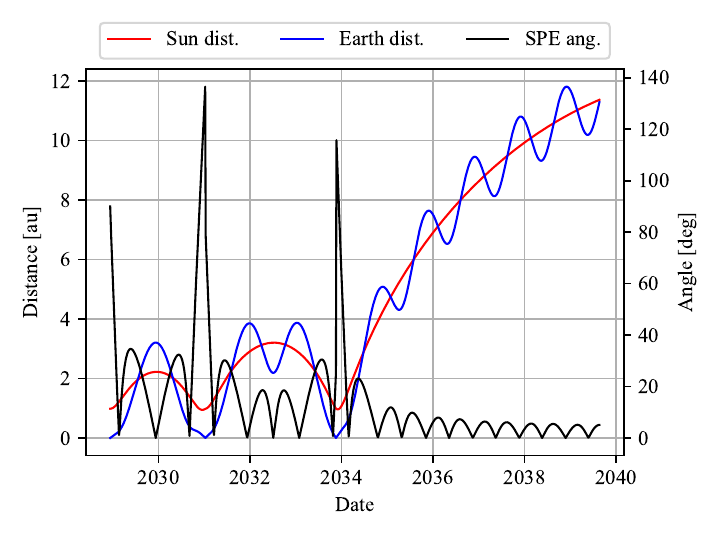}
		\subcaption{Geometry}
		\label{fig:fb_eeea_geom}
	\end{minipage}
	\caption{Flyby trajectory with the EEEA sequence (launch on December 10, 2028).}
	\label{fig:fb_eeea}
\end{figure}

The actual trajectories and itineraries of two distinct cases, EEMEJA (No.~1) and EEEA (No.~6), at the first opportunity are shown in Figs.~\ref{fig:fb_eemeja}--\ref{fig:fb_eeea} and Tables~\ref{tab:fb_eemeja}--\ref{tab:fb_eeea}. Note that the SPE angle stands for Sun--Probe--Earth angle which is an important factor in designing the communication system. Figure~\ref{fig:fb_eemeja} confirms that, in the EEMEJA case, the \astname flyby occurs at a high latitude in the J2000EC system, with a solar distance of 9.2~au and an Earth distance of 9.8~au. In the EEEA sequence, the spacecraft reaches \astname on the ecliptic plane at a solar distance of 11.3~au and an Earth distance of 11.4~au. The maximum solar and Earth distances are smaller in the EEMEJA case compared to EEEA, which is advantageous from the perspective of spacecraft systems. Conversely, the EEEA sequence results in a smaller relative velocity at the \astname encounter, which can alleviate the requirements for the observation system.

\begin{table}
	\centering
	\caption{Itinerary for the flyby mission with EEMEJA sequence}
	\label{tab:fb_eemeja}
	\begin{tabular}{llccc}
		\hline
		Event  &  Epoch  &  $V_\infty$ [km/s]  &  Flyby altitude [km]  &  $\dv$ [m/s] \\
		\hline
		Earth  &  2033-Nov-27  &  5.10  &  --  &  0 \\
		DSM-0  &  2034-Oct-25  &  --  &  --  &  61.87 \\
		Earth  &  2035-Nov-03  &  5.66  &  20293.61  &  0  \\
		Mars  &  2038-Jun-10  &  9.56  &  1135.62  &  0  \\
		Earth  &  2039-Dec-18  &  11.31  &  638.17  &  0 \\
		Jupiter  &  2041-Oct-18  &  8.45  &  650639.49  &  29.36  \\
		\astname  &  2044-Nov-17  &  10.02  &  --  &  0 \\
		\hline
	\end{tabular}
\end{table}

\begin{table}
	\centering
	\caption{Itinerary for the flyby mission with EEEA sequence}
	\label{tab:fb_eeea}
	\begin{tabular}{llccc}
		\hline
		Event  &  Epoch  &  $V_\infty$ [km/s]  &  Flyby altitude [km]  &  $\dv$ [m/s] \\
		\hline
		Earth  &  2028-Dec-10  & 5.10  &  --  & 0 \\
		DSM-0  &  2029-Dec-15  &  --  &  --  &  222.78  \\
		Earth  &  2031-Jan-09  &  6.95  &  9588.04  &  0 \\
		DSM-1  &  2032-Jun-08  &  --  &  --  &  542.97  \\
		Earth  &  2033-Nov-16  &  11.77  &  637.82  &  0  \\
		\astname  &  2039-Aug-21  &  6.97  &  --  &  0 \\
		\hline
	\end{tabular}
\end{table}

\subsection{Rendezvous Mission}\label{sec34}
Similarly to the flyby mission design presented in Sec.~\ref{sec33}, potential flyby sequences were generated first. In a rendezvous mission, an SGA or a JGA is essential to align with \astname's orbital plane. Including the SGA or JGA that occurs during the final interplanetary flyby, up to five intermediate bodies were considered in the SGA case and four bodies in the JGA case. Table~\ref{tab:rv_bodies} shows a list of bodies for each intermediate flyby. The first flyby is limited to inner planets, i.e., Earth, Venus, and Mars, while the final flyby is constrained to either Saturn or Jupiter. By excluding redundant and unrealistic sequences in a manner similar to that described in Sec.~\ref{sec33}, 190~sequences were generated. The box bounds on the departure $V_\infty$ were set to the same value as the flyby mission design, specifically [1, 5.1]~km/s. As shown in Fig.~\ref{fig:geom}, an SGA on February 11, 2045, and a JGA on August 28, 2042, will be effective for raising the inclination. Therefore, the box bounds on the departure epoch $t_0$ was established as [2030-Jan-01, 2040-Jan-01]. Given that a rendezvous mission requires more time than a flyby mission, it is impractical to consider later opportunities; hence, only the [2030-Jan-01, 2040-Jan-01] window is addressed in the rendezvous mission design. The optimization settings remain consistent with those of the flyby mission design (see also Table~\ref{tab:optset}). The upper bound on the total TOF was set to 9000~days in the SGA case and 5500~days in the JGA case.

\begin{table}
	\centering
	\caption{Possible combinations of intermediate bodies for rendezvous missions}
	\label{tab:rv_bodies}
	\begin{tabular}{ccc}
		\hline
		Body  &  Integer code (SGA case)  &  Integer code (JGA case)  \\
		\hline
		P$_1$  &  [0, 1, 2]  &  [0, 1, 2]  \\
		P$_2$  &  [-1, 0, 1, 2, 3]  &  [-1, 0, 1, 2, 3]  \\
		P$_3$  &  [-1, 0, 1, 2, 3]  &  [-1, 0, 1, 2, 3]  \\
		P$_4$  &  [-1, 0, 1, 2, 3]  &  [3]  \\
		P$_5$  &  [4]  &  [-1] \\
		\hline
	\end{tabular}
\end{table}

Tables~\ref{tab:opt_rv1} and \ref{tab:opt_rv2} show the optimization results, confirming that trajectories via Saturn yield a smaller $\dv$ but a longer flight time. Notably, a rendezvous can be achieved with a $\dv$ of approximately 2~km/s, which is a suitable value for a chemical propulsion mission. However, it is important to note that a flight time of 24.6~years poses significant challenges. In contrast, the journey via Jupiter requires a greater $\dv$ but can be completed in about 13.4~years. Given that exploration missions to outer planets, such as the Uranus Orbiter \& Probe \cite{bib:uop}, typically necessitate similar flight durations, this finding is both realistic and appealing.

\begin{table}
	\centering
	\caption{Top 10 sequences for rendezvous missions with an SGA}
	\label{tab:opt_rv1}
	\begin{tabular}{clllccc}
		\hline
		No.  &  Sequence  &  Departure  &  Arrival  &  $\dv$ [m/s]  &  P-FB  &  DSM  \\
		\hline
		1  &  EMEEJSA  &  2033-Aug-03  &  2058-Mar-23  &  1977.06  &  True  &  False \\
		2  &  EVEEJSA  &  2034-Mar-28  &  2058-Nov-17  &  2030.86  &  True  &  True  \\
		3  &  EEESA  &  2036-Feb-02  &  2060-Sep-23  &  2309.93  &  False  &  True \\
		4  &  EMEESA  &  2035-Sep-02  &  2060-Apr-23  &  2453.56  &  True  &  True \\
		5  &  EVEESA  &  2034-Jun-12  &  2059-Feb-01   &  2470.95  &  True  &  True \\
		6  &  EEMESA  &  2033-Nov-26  &  2058-Jul-18  &  2721.44  &  False  &  True \\
		7  &  EEEJSA  &  2034-Sep-24  &  2059-May-16  &  2881.27  &  True  &  True \\
		8  &  EEMEJSA  &  2032-Mar-12  &  2056-Oct-31  &  2905.35  &  True  &  True \\
		9  &  EEVEJSA  &  2034-Aug-17  &  2059-Apr-08  &  2958.68  &  True  &  True \\
		10  &  EMVEJSA  &  2035-Jul-31  &  2060-Mar-21  &  3022.94  &  True  &  True \\
		\hline
	\end{tabular}
\end{table}

\begin{table}
	\centering
	\caption{Top 10 sequences for rendezvous missions with a JGA}
	\label{tab:opt_rv2}
	\begin{tabular}{clllccc}
		\hline
		No.  &  Sequence  &  Departure  &  Arrival  &  $\dv$ [m/s]  &  P-FB  &  DSM  \\
		\hline
		1  &  EMEEJA  &  2035-Sep-18  &  2049-Feb-24  &  2716.56  &  False  &  True \\
		2  &  EVEEJA  &  2035-Aug-15  &  2049-Mar-28  &  2757.19  &  False  &  True  \\
		3  &  EVVEJA  &  2037-Sep-12  &  2049-May-23  &  3015.06  &  False  &  True \\
		4  &  EEEJA  &  2035-Dec-24  &  2049-Mar-04  &  3141.55  &  False  &  True  \\
		5  &  EEJA  &  2039-Jan-27  &  2053-Feb-11  &  3611.12  &  False  &  True \\
		6  &  EMMEJA  &  2033-Oct-10  &  2048-Oct-31  &  3788.53  &  False  &  True \\
		7  &  EMEJA  &  2037-Sep-09  &  2052-Jun-17  &  3991.34  &  False  &  True \\
		8  &  EEMEJA  &  2035-Sep-23  &  2049-Apr-21  &  4001.36  &  False  &  True   \\
		9  &  EVEJA  &  2036-Feb-19  &  2051-Mar-09  &  4081.57 &  True  &  True   \\
		10 &  EMVEJA  &  2031-Apr-28  &  2046-May-19  &  4321.54  &  False  &  True \\
		\hline
	\end{tabular}
\end{table}

\begin{figure}
	\centering
	\begin{minipage}{0.49\linewidth}
		\includegraphics[width=\linewidth]{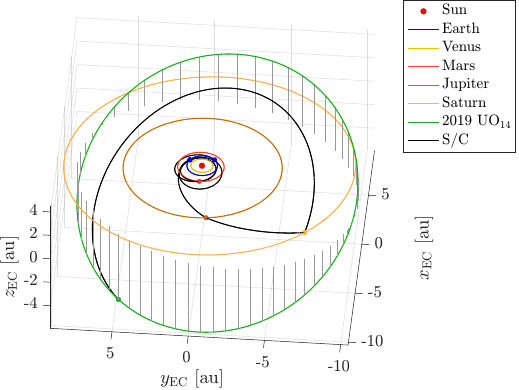}
		\subcaption{Trajectory}
		\label{fig:rv_emeejsa_traj}
	\end{minipage}
	\begin{minipage}{0.49\linewidth}
		\includegraphics[width=\linewidth]{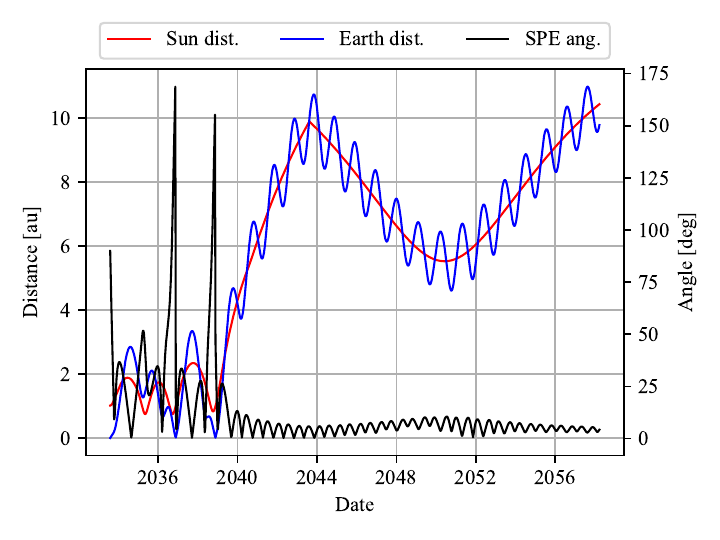}
		\subcaption{Geometry}
		\label{fig:rv_emeejsa_geom}
	\end{minipage}
	\caption{Rendezvous trajectory with EMEEJSA sequence (launch on August 3, 2033).}
	\label{fig:rv_emeejsa}
\end{figure}

\begin{table}
	\centering
	\caption{Itinerary for the rendezvous mission with EMEEJSA sequence}
	\label{tab:rv_emeejsa}
	\begin{tabular}{llccc}
		\hline
		Event  &  Epoch  &  $V_\infty$ [km/s]  &  Flyby altitude [km]  &  $\dv$ [m/s] \\
		\hline
		Earth  &  2033-Aug-03  &  4.50  &  --  &  0 \\
		Mars  &  2034-Nov-02  &  5.18  &  1105.85  &  0 \\
		Earth  &  2036-Nov-19  &  11.89  &  13515.88  &  0 \\
		Earth  &  2038-Nov-20  &  11.88  &  749.71  &  0 \\
		Jupiter  &  2040-Aug-14  &  8.98  &  4.22$\times$10$^6$  &  0 \\
		Saturn  &  2043-Aug-14  &  5.69  &  254014.43  &  154.33 \\
		\astname  &  2058-Mar-23  &  1.82  &  --  &  1822.46 \\
		\hline
	\end{tabular}
\end{table}

\begin{figure}
	\centering
	\begin{minipage}{0.49\linewidth}
		\includegraphics[width=\linewidth]{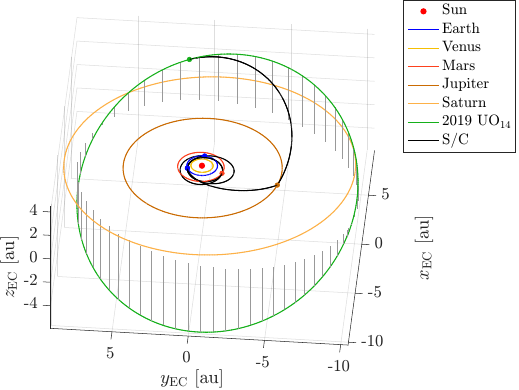}
		\subcaption{Trajectory}
		\label{fig:rv_emeeja_traj}
	\end{minipage}
	\begin{minipage}{0.49\linewidth}
		\includegraphics[width=\linewidth]{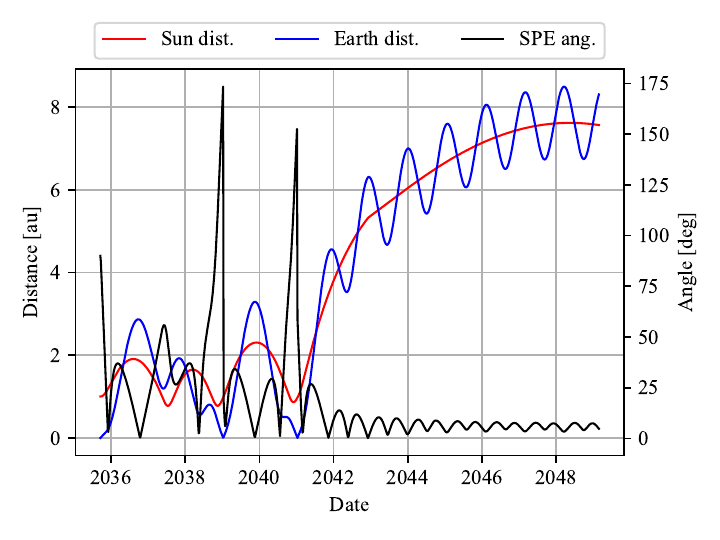}
		\subcaption{Geometry}
		\label{fig:rv_emeeja_geom}
	\end{minipage}
	\caption{Rendezvous trajectory with EMEEJA sequence (launch on September 18, 2035).}
	\label{fig:rv_emeeja}
\end{figure}

Two representative cases from the rendezvous mission design results are shown in Figs.~\ref{fig:rv_emeejsa}--\ref{fig:rv_emeeja} and Tables~\ref{tab:rv_emeejsa}--\ref{tab:rv_emeeja}. In the EMEEJSA sequence, both the maximum solar distance and the maximum Earth distance exceed 10~au due to the arrival around the aphelion of \astname, which presents another drawback for this sequence. In contrast, the maximum solar distance and the maximum Earth distance in the EMEEJA sequence are 7.6~au and 8.5~au, respectively, as the spacecraft arrives at \astname near its perihelion. This results in a substantial reduction in the spacecraft system requirements compared to the EMEEJSA case.

\begin{table}
	\centering
	\caption{Itinerary for the rendezvous mission with EMEEJA sequence}
	\label{tab:rv_emeeja}
	\begin{tabular}{llccc}
		\hline
		Event  &  Epoch  &  $V_\infty$ [km/s]  &  Flyby altitude [km]  &  $\dv$ [m/s] \\
		\hline
		Earth  &  2035-Sep-18  &  4.40  &  --  &  0 \\
		Mars  &  2037-Feb-04  &  5.36  &  767.77  &  0 \\
		Earth  &  2039-Jan-10  &  10.22  &  13288.37  &  0 \\
		Earth  &  2041-Jan-09  &  10.23  &  1631.59  &  0 \\
		Jupiter  &  2042-Dec-08  &  8.05  &  1.73$\times$10$^6$  &  0 \\
		\astname  &  2049-Feb-24  &  2.72  &  --  &  2716.44 \\
		\hline
	\end{tabular}
\end{table}

\section{Low-Thrust Trajectories}\label{sec4}
The analyses in the previous section demonstrated that a flyby mission can be accomplished with nearly zero $\dv$, whereas a rendezvous mission necessitates a significant $\dv$. Notably, the $\dv$s must be performed in the outer solar system, particularly during the powered flyby of Saturn or at the \astname arrival (see Tables~\ref{tab:rv_emeejsa} and \ref{tab:rv_emeeja}). In other words, $\dv$ maneuvers are required only in the final trajectory segment (i.e., Saturn--\astname or Jupiter--\astname). Due to the slow orbital motion in the outer solar system, low-thrust propulsion is expected to be effective there. Therefore, this section investigates low-thrust trajectories for rendezvousing with \astname. In particular, the EMEEJA sequence, which offers a short flight time and a relatively large $\dv$, is examined.

\subsection{Trajectory Model}\label{sec41}
Provided that $\dv$s are primarily required in the outer solar system and that low-thrust propulsion is less effective in the inner solar system due to a short time constant, it is assumed that low-thrust propulsion is used only during the final arc (i.e., from Jupiter to \astname). In other words, the final arc of the ballistic trajectory obtained in Sec.~\ref{sec34} is transformed into a low-thrust arc. To achieve this, a combined ballistic--low-thrust trajectory optimization problem is formulated.

The ballistic phase (i.e., from launch to Jupiter) follows the same formulation as described in Sec.~\ref{sec31}. Specifically, the decision vector is obtained by simply replacing $N$ with $N - 1$ as follows:
\begin{align}
		\bm{X}_\mathrm{bal} = [t_0, \bm{Y}_1, \cdots, \bm{Y}_{N - 1}]
\end{align}
Although impulsive maneuvers are possible if the spacecraft is equipped with both chemical and electric engines, it is assumed in this study that only low-thrust propulsion is used. To achieve this, the $\dv$s in the ballistic arcs are handled not as objectives but as constraints; that is, the following equality constraints must hold true:
\begin{align}
	\begin{aligned}
		\dv_{\mathrm{DSM}, i} &= 0 & \text{ for } N = 1, \cdots, N - 1 \\
		\dv_{\mathrm{P-FB}, i + 1} &= 0 & \text{ for } N = 1, \cdots, N - 1
	\end{aligned}
	\label{eq:ec_dsm}
\end{align}

As for the low-thrust arc, the direct collocation method \cite{bib:dcnlp}, which has been extensively applied in various low-thrust trajectory design problems, is employed. To gain a comprehensive understanding of low-thrust missions, a normalized dynamics model is used. The equation of motion is written as
\begin{align}
	\begin{aligned}
		\dot{\bm{r}} &= \bm{v} \\
		\dot{\bm{v}} &= -\frac{\mu}{\| \bm{r} \|^3} \bm{r} + \frac{T}{M}\bm{u}
	\end{aligned}
	\label{eq:eom1}
\end{align}
where $\bm{r}$ is the position vector, $\bm{v}$ is the velocity vector, $\mu$ is the heliocentric gravitational parameter, $T$ is the maximum thrust, $M$ is the spacecraft mass, and $\bm{u}$ is the throttle vector. The spacecraft mass is dominated by the following Tsiolkovsky rocket equation:
\begin{align}
	\dot{M} = -\frac{T}{g \Isp} \| \bm{u} \|
	\label{eq:rocket1}
\end{align}
where $g$ is the standard gravitational acceleration and $\Isp$ is the specific impulse. Let $m = M / M_0 \in [0, 1]$ represent the mass ratio, which indicates the mass fraction relative to the initial mass $M_0$, and let $a_0 = T / M_0$ denote the characteristic acceleration. Consequently, Eqs~(\ref{eq:eom1}) and (\ref{eq:rocket1}) can be reformulated as
\begin{align}
	&
	\begin{aligned}
		\dot{\bm{r}} &= \bm{v} \\
		\dot{\bm{v}} &= -\frac{\mu}{\| \bm{r} \|^3} \bm{r} + \frac{a_0}{m}\bm{u}
	\end{aligned}
	\label{eq:eom2}
	\\
	&\dot{m} = -\frac{a_0}{g \Isp} \| \bm{u} \|
	\label{eq:rocket2}
\end{align}
Thus, the mission-specific parameters are encapsulated in $a_0$ and $\Isp$. By using Eqs.~(\ref{eq:eom2}) and (\ref{eq:rocket2}), the low-thrust trajectories can be analyzed without the need to define the thrust level and the spacecraft mass.

In the direct collocation method, the trajectory is discretized into multiple nodes. Letting $L$ be the number of discrete nodes, the decision vector in the low-thrust arc is given by
\begin{align}
	\bm{X}_\mathrm{LT} = [\tau_N, \bm{r}_1, \bm{v}_1, m_1, \bm{u}_1, \cdots, \bm{r}_L, \bm{v}_L, m_L, \bm{u}_L]
\end{align}
Let the term ``segment'' refer to the region between two adjacent nodes. The state history and control history in each segment are interpolated using cubic polynomials and a linear function, respectively. The following equality constraint is then imposed at the center of each segment:
\begin{align}
	\hat{\dot{\bm{x}}}_k = h(\hat{\bm{x}}_k, \hat{\bm{u}}_k) \quad \text{for } k = 1, \cdots, L - 1
	\label{eq:ec_dyn}
\end{align}
where $\hat{(\cdot)}$ denotes the center value of the interpolated curve, $\bm{x} = [\bm{r}, \bm{v}, m]$ is the extended state vector of the spacecraft, and $h(\bm{x}, \bm{u})$ yields the time derivative of $\bm{x}$ subject to the control $\bm{u}$. Note that $h(\bm{x}, \bm{u})$ is identical to the right-hand sides of Eqs.~(\ref{eq:eom2}) and (\ref{eq:rocket2}). The boundary conditions are equivalent to equality constraints, which are given by
\begin{align}
	\begin{aligned}
		\bm{r}_1 &= \bm{r}_{p, N} \\
		m_1 &= 1 \\
		\bm{r}_L &= \bm{r}_{p, N + 1} \\
		\bm{v}_L &= \bm{v}_{p, N + 1}
	\end{aligned}
	\label{eq:ec_bc}
\end{align}
Note that the boundary condition for the initial velocity $\bm{v}_1$ is incorporated in the P-FB constraint, i.e., in Eq.~(\ref{eq:ec_dsm}). From its definition, the throttle vector must satisfy the following inequality constraint:
\begin{align}
	\bm{u}_k \cdot \bm{u}_k \le 1 \quad \text{for } k = 1, \cdots, L
	\label{eq:ic_thr}
\end{align}
Additionally, the maximum TOF constraint, which was incorporated as a penalty factor in Sec.~\ref{sec31}, is handled exactly as an inequality constraint as follows:
\begin{align}
	\sum_{i = 1}^{N} \tau_i - \tau_\mathrm{max} \le 0
	\label{eq:ic_tof}
\end{align}
The objective function is the $\dv$ due to low-thrust propulsion, which is given by
\begin{align}
	\dv_\mathrm{LT} = g \Isp \ln \frac{m_1}{m_L}
\end{align}
Thus, letting $\bm{X} = [\bm{X}_\mathrm{bal}, \bm{X}_\mathrm{LT}]$ be the new decision vector, the optimization problem can be formulated as follows:
\begin{gather}
	\begin{gathered}
		\min_{\bm{X}} \dv_\mathrm{LT}  \\
		\textrm{subject to} \\
		\text{Equality constraints: Eqs.~(\ref{eq:ec_dsm}), (\ref{eq:ec_dyn}), and (\ref{eq:ec_bc})} \\
		\text{Inequality constraints: Eqs.~(\ref{eq:ic_thr}) and (\ref{eq:ic_tof})}
	\end{gathered}
	\label{eq:opt_lt}
\end{gather}

\subsection{Solution Method}\label{sec42}
The optimization problem defined by Eq.~(\ref{eq:opt_lt}) is solved using the SQP algorithm. The initial guess for the ballistic term $\bm{X}_\mathrm{bal}$ is derived from the solutions obtained in Sec.~\ref{sec3}. For the low-thrust term $\bm{X}_\mathrm{LT}$, the initial guesses for the positions $\bm{r}_1, \cdots, \bm{r}_L$ and the velocities $\bm{v}_1, \cdots, \bm{v}_L$ are extracted from the final leg of the ballistic trajectory. A series of random values within the range of $[-1, 1]$ is used for the initial guess of the control inputs $\bm{u}_1, \cdots, \bm{u}_L$. Beginning with $m_1 = 1$, the initial guess for the mass parameter $m_k$ is determined such that the following equation is satisfied:
\begin{align}
	m_{k + 1} = m_k - \frac{a_0}{g \Isp} \| \bm{u}_k \| \Delta \tau
\end{align}
where $\Delta \tau = \tau_N / (L - 1)$ is the flight time for each segment.

Because the initial guess includes random values, convergence is not guaranteed. Therefore, the optimization process is repeated multiple times using different random seeds until no further improvement is observed. Test cases have confirmed that the optimization converges reliably and five trials are sufficient to identify the optimal solution.

\subsection{Analysis and Results}\label{sec43}
\begin{figure}[!b]
	\centering
	\includegraphics[width=0.5\linewidth]{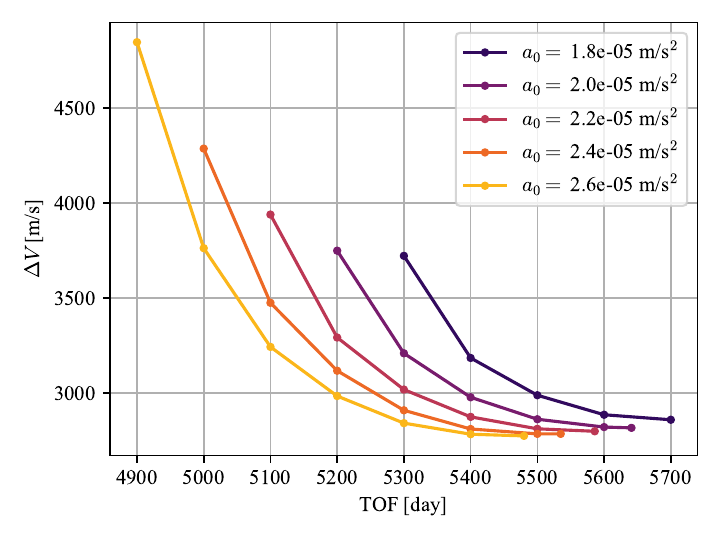}
	\caption{Trade-off for low-thrust trajectories between $\dv$ and TOF.}
	\label{fig:pareto}
\end{figure}

A parametric study was conducted to examine low-thrust trajectories to \astname. Specifically, the optimization problem defined in Sec.~\ref{sec41} was solved with various characteristic accelerations. The maximum TOF was also varied to evaluate the trade-off between $\dv$ and TOF\@. The $\Isp$ was fixed to 3000~s because it has minimal impact on the optimization results, except for the mass history. Several trials confirmed that feasible solutions exist at an acceleration level of around $2 \times 10^{-5}$~m/s$^2$ and a TOF of around 5300~days. Figure~\ref{fig:pareto} shows the trade-off between $\dv$ and TOF for various $a_0$, demonstrating that $\dv$ and TOF are in a competing relationship. The Pareto front shifts in an optimal direction with increasing acceleration levels. It is challenging to reduce the TOF to less than 5000~days, while extending it beyond 5500~days proves to be less effective.

\begin{figure}
	\centering
	\includegraphics[width=0.5\linewidth]{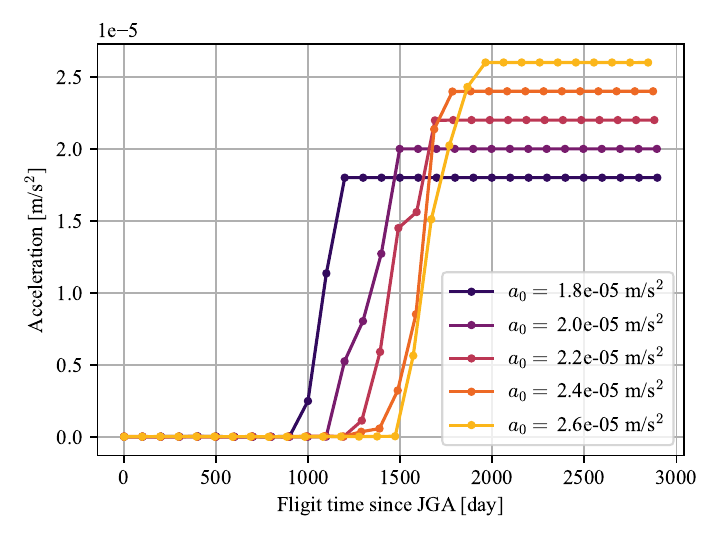}
	\caption{Thrust profile in the $\tau_\mathrm{max}=5500$~days case}
	\label{fig:thrust}
\end{figure}

Figure~\ref{fig:thrust} shows the thrust history for various $a_0$ with a fixed TOF of 5500~days. Similar to typical cases, the thrust profile displays a Bang--Bang structure. The $\dv$ maneuver is primarily required before arrival. As the acceleration increases, the duration of the thrust duty decreases. This figure reinforces the notion that the low-thrust trajectories are feasible with an acceleration level of around $2 \times 10^{-5}$~m/s$^2$; specifically, the thrust duty will exceed the allowable flight time if the acceleration is smaller than this value.

\begin{figure}[!b]
	\centering
	\begin{minipage}{0.49\linewidth}
		\includegraphics[width=\linewidth]{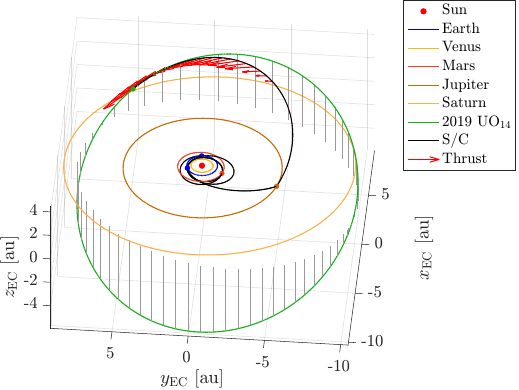}
		\subcaption{J2000EC}
		\label{fig:lt_emeeja_ec}
	\end{minipage}
	\begin{minipage}{0.49\linewidth}
		\includegraphics[width=\linewidth]{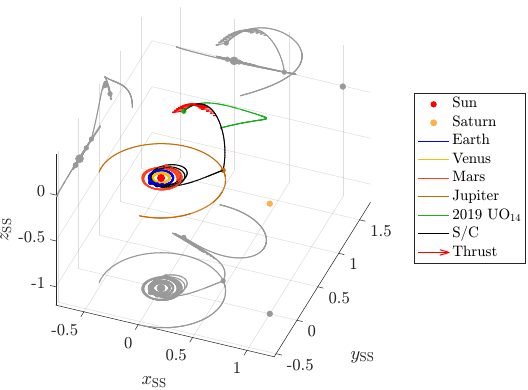}
		\subcaption{Sun--Saturn frame}
		\label{fig:lt_emeeja_ss}
	\end{minipage}
	\caption{Low-thrust trajectory with $a_0=2.0 \times 10^{-5}$~m/s$^2$ and $\Isp = 3000$~s.}
	\label{fig:lt_emeeja}
\end{figure}

Figure~\ref{fig:lt_emeeja} illustrates an optimal low-thrust trajectory with $a_0 = 2 \times 10^{-5}$~m/s$^2$ and $\tau_\mathrm{max} = 5500$~days. Figure~\ref{fig:lt_emeeja_ec} confirms that the shape of the low-thrust trajectory is nearly identical to that of the ballistic trajectory (see Fig.~\ref{fig:rv_emeeja_traj}). Figure~\ref{fig:lt_emeeja_ss} indicates that the spacecraft travels to the Sun--Saturn $L_4$ point following the Jupiter flyby. As discussed in Sec.~\ref{sec2}, the orbital motion of \astname is subject to significant perturbations around the Sun--Saturn $L_4$ point. Due to these perturbations, the trajectory of \astname approaches that of Jupiter around 2043 when observed in the SS frame. This is why the route via Jupiter can be a shortcut to the location of \astname. Given that the maximum solar distance is approximately 7.5~au, the sequence involving a Jupiter flyby is beneficial from the perspective of solar electric propulsion; in contrast, the electric engines must operate at a solar distance exceeding 10~au in sequences involving a Saturn flyby.

\subsection{Discussion on Required Acceleration Levels}\label{sec44}
The previous analyses demonstrated that low-thrust missions to \astname are possible without the necessity of operating electric thrusters at a Saturn distance. Specifically, the maximum solar distance is 7.5~au if a gravity assist from Jupiter is employed. However, ensuring adequate solar power at a distance of 7.5~au remains a significant challenge. In this section, the possibility of low-thrust missions to \astname is briefly discussed by comparing them with actual missions that have been launched or were planned.

\begin{figure}
	\centering
	\includegraphics[width=0.5\linewidth]{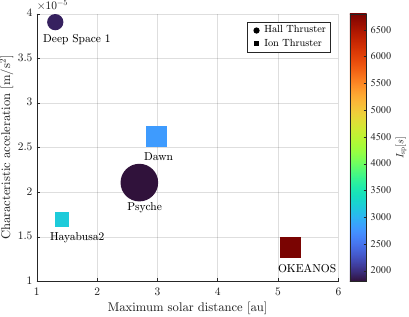}
	\caption{Acceleration levels of the actual mission. The marker size corresponds to the launch mass of the spacecraft.}
	\label{fig:acceleration}
\end{figure}

Figure~\ref{fig:acceleration} shows the characteristic acceleration plotted against the maximum solar distance for actual missions: Deep Space~1 \cite{bib:ds1}, Dawn \cite{bib:dawn}, Hayabusa2 \cite{bib:mu10}, Psyche \cite{bib:psyche}, and OKEANOS \cite{bib:okeanos1}. Considering that electric thrusters must operate in regions far from the Sun for the \astname exploration mission, the characteristic accelerations depicted in Fig.~\ref{fig:acceleration} were evaluated in the low-power mode of the corresponding missions. This figure confirms that the acceleration levels of the actual missions are on the order of $10^{-5}$~m/s$^2$, which is compatible with the requirements for the \astname exploration mission. The characteristic acceleration decreases with increasing solar distance. The state-of-the-art technologies for outer solar system exploration are represented by the OKEANOS mission, which operates its electric thrusters at a solar distance of 5.2~au. To potentially achieve a sample return from the Jupiter Trojans, the OKEANOS mission requires a significantly large $\dv$ of over 5500~m/s. For this reason, OKEANOS is equipped with high-$\Isp$ ion engines. In contrast, the required $\dv$ for \astname explorations is approximately 3000~m/s, which is considerably smaller than that of the OKEANOS mission. In fact, the propellant mass fraction for \astname exploration is 10~\% of the wet mass when using a thruster with an $\Isp$ of 3000~s. Therefore, higher acceleration levels may be achievable for outer solar system explorations if the specific impulse is reduced from OKEANOS's ion engines. Thus, developing an electric propulsion system capable of producing an acceleration of $2.0 \times 10^{-5}$~m/s$^2$ is a realistic scenario. Although several technical challenges are still present, low-thrust missions to \astname could be realized using existing technologies.

\section{Conclusions}\label{sec5}
In this paper, mission analysis to explore the first-ever Saturn Trojan \astname was conducted. Given that \astname has a large orbital inclination and a long orbital period of over 30~years, opportunities for this object to cross the ecliptic plane occur approximately every 15~years. This characteristic may significantly restrict launch windows; therefore, a comprehensive investigation of possible scenarios for reaching \astname was undertaken.

First, assuming the use of chemical engines, ballistic flight trajectories were examined. A meta-heuristic global optimization algorithm was developed to solve a trajectory optimization problem involving both DSMs and P-FBs. All potential flyby sequences were analyzed in a parallelized manner. The results indicated that an encounter with \astname is possible outside the ecliptic plane by utilizing a gravity assist from Jupiter, thereby relaxing the requirements for the launch windows. For the flyby option, two promising solutions were identified through the global optimization process: a departure in November 2033 and another in March 2039. Both scenarios involve a JGA at the end of intermediate flybys, resulting in a $\dv$ of less than 100~m/s and a flight time of 11~years. Other windows that may be used as buckup opportunities require a $\dv$ of less than 1~km/s. For the rendezvous option, two scenarios were analyzed, specifically the trajectory via Saturn and the one via Jupiter. Since \astname belongs to the Sun--Saturn dynamical system, a flight through Saturn is the most likely scenario. As expected, this route involving an SGA offers the minimal $\dv$ of 1977~m/s with a departure in August 2033. However, due to the slow orbital motion in the Saturn region, a flight time of 24.6~years is required. In contrast, a JGA provides a shortcut to \astname, resulting in a $\dv$ of 2717~m/s and a flight time of 13.4~years with a departure in September 2035. Given that other missions to outer planets, such as Voyager~1 \& 2 and the Uranus Orbiter and Probe, require a similar flight time, this solution is both realistic and appealing.

In the rendezvous missions, $\dv$ maneuvers are required primarily in the outer solar system. In this region, low-thrust propulsion is effective because there is sufficient time available to accelerate the spacecraft. To gain a comprehensive understanding of low-thrust trajectories, a normalized model was developed, eliminating the need for specific values of spacecraft mass and thrust magnitude. A parametric study on characteristic acceleration demonstrated that acceleration levels ranging from $1.8 \times 10^{-5}$~m/s$^2$ to $2.6 \times 10^{-5}$~m/s$^2$ are preferred. Notably, the low-thrust mission can be accomplished within a time frame nearly equivalent to that of chemical propulsion missions.

As described previously, \astname's orbit presents significant challenges to reach from an engineering point of view. However, the global trajectory optimization technique introduced in this study has yielded reliable solutions that are feasible with existing technologies. The best launch windows occur between 2033 and 2035, indicating that the late 2020s will be the ideal time to initiate exploration missions. Although classified as a Saturn Trojan, \astname's orbit largely oscillates around the Sun--Saturn $L_4$ point. In the late 2040s, which is the expected arrival period, \astname will approach Jupiter's orbit. This means that the spacecraft system does not need to be designed for the Saturn distance, further enhancing the reliability of the presented solution. As highlighted in the paper on the discovery of \astname, this object possesses intriguing and enigmatic characteristics. Once the exploration of \astname is realized using the trajectories identified in the present study, it is expected that our understanding of the solar system's formation processes will be significantly advanced.

\section*{Appendix: Generation of Initial Bounds on the Solution Space}\label{secA}
\renewcommand{\theequation}{A.\arabic{equation}}
\setcounter{equation}{0}

A method for determining the initial bounds on the decision variables in the global optimization of ballistic trajectories is presented here. As described in Sec.~\ref{sec31}, the decision variables include the departure epoch, the infinity velocity at the departure and intermediate bodies, the TOF, and the DSM index. Among these variables, the $V_\infty$ direction and the DSM index are constrained as follows:
\begin{align}
	\begin{aligned}
		\alpha & \in [-\pi, \pi] \\
		\beta & \in \left[ -\frac{\pi}{2}, \frac{\pi}{2} \right] \\
		\eta & \in [0, 1]
	\end{aligned}
\end{align}
The bounds on the departure epoch $t_0$ must be specified by the users. Therefore, the $V_\infty$ magnitude and the TOF are the remaining parameters that need to be bounded.

The departure $V_\infty$ magnitude at Earth is directly related to the performance of the launch vehicle; therefore, it must also be specified by the user. In this study, the bounds on the departure $V_\infty$ was set to [1, 5.1]~km/s. For the subsequent flybys, the lower bound was uniformly established at 2~km/s. This value is adequate to prevent low-velocity flybys. In addition, flybys with a $V_\infty$ smaller than this threshold rarely occur in actual trajectories. The upper bound on the $V_\infty$ magnitude for the inner planets (i.e., Earth, Venus, and Mars) was set to 1.1 times their escape velocity, while 9~km/s was used for the outer planets (i.e., Jupiter and Saturn).

The bounds on the TOF for each trajectory leg were determined based on the combination of the initial and terminal bodies. Let $P_s$ and $P_H$ represent the synodic period and the Hohmann transfer time between the two bodies, respectively. If the two bodies consist solely of inner planets, the bounds were determined as follows:
\begin{align}
	\tau \in [30 \text{ days}, P_s + P_H]
\end{align}
Otherwise, i.e., if the outer planets or \astname are associated, the bounds were set to
\begin{align}
	\tau \in [0.3 P_H, 1.3 P_H]
\end{align}
Thus, users need to specify only the epoch and the $V_\infty$ magnitude at launch. The bounds on other parameters are automatically determined based on the sequence of bodies, allowing for the fully automated design of MPGA-1DSM trajectories.

\bibliographystyle{unsrt}  
\bibliography{references}

\begin{thebibliography}{10}

\bibitem{bib:nice1}
Kleomeris Tsiganis, Rodney Gomes, Alessandro Morbidelli, and Hal~F Levison.
\newblock Origin of the orbital architecture of the giant planets of the {Solar} {System}.
\newblock {\em Nature}, 435(7041):459--461, 2005.

\bibitem{bib:nice2}
Alessandro Morbidelli, Harold~F Levison, Kleomenis Tsiganis, and Rodney Gomes.
\newblock Chaotic capture of {Jupiter's} {Trojan} asteroids in the early {Solar} {System}.
\newblock {\em Nature}, 435(7041):462--465, 2005.

\bibitem{bib:tack}
Kevin~J Walsh, Alessandro Morbidelli, Sean~N Raymond, David~P O'Brien, and Avi~M Mandell.
\newblock A low mass for {Mars} from {Jupiter's} early gas-driven migration.
\newblock {\em Nature}, 475(7355):206--209, 2011.

\bibitem{bib:et}
Man-To Hui, Paul~A Wiegert, David~J Tholen, and Dora F{\"o}hring.
\newblock The second {Earth} {Trojan} 2020 {XL$_5$}.
\newblock {\em The Astrophysical Journal Letters}, 922(2):L25, 2021.

\bibitem{bib:uo14}
Man-To Hui, Paul~A Wiegert, Robert Weryk, Marco Micheli, David~J Tholen, Sam Deen, Andrew~J Walker, and Richard Wainscoat.
\newblock {2019 UO$_{14}$}: A transient {Trojan} of {Saturn}.
\newblock {\em The Astrophysical Journal Letters}, 975(1):L3, 2024.

\bibitem{bib:lucy}
Harold~F Levison, Simone Marchi, Keith Noll, Catherine Olkin, Thomas~S Statler, Lucy~Science Team, et~al.
\newblock {NASA}'s {Lucy} mission to the {Trojan} asteroids.
\newblock In {\em 2021 IEEE Aerospace Conference (50100)}, pages 1--10, Big Sky, MT, USA, 2021. IEEE.

\bibitem{bib:okeanos1}
Osamu Mori, Jun Matsumoto, Toshihiro Chujo, Masanori Matsushita, Hideki Kato, Takanao Saiki, Yuichi Tsuda, Jun'ichiro Kawaguchi, Fuyuto Terui, Yuya Mimasu, et~al.
\newblock Solar power sail mission of {OKEANOS}.
\newblock {\em Astrodynamics}, 4(3):233--248, 2020.

\bibitem{bib:okeanos2}
Yuki Takao, Osamu Mori, Jun Matsumoto, Toshihiro Chujo, Shota Kikuchi, Yoko Kebukawa, Motoo Ito, Tatsuaki Okada, Jun Aoki, Kazuhiko Yamada, et~al.
\newblock Sample return system of {OKEANOS}---the solar power sail for jupiter trojan exploration.
\newblock {\em Acta Astronautica}, 213:121--137, 2023.

\bibitem{bib:et_exp}
Hanlun Lei, Bo~Xu, and Lei Zhang.
\newblock Trajectory design for a rendezvous mission to {Earth's Trojan} asteroid {2010 TK7}.
\newblock {\em Advances in Space Research}, 60(11):2505--2517, 2017.

\bibitem{bib:lucy2}
Jacob~A Englander, Donald~H Ellison, Ken Williams, James McAdams, Jeremy~M Knittel, Brian Sutter, Chelsea Welch, Dale Stanbridge, and Kevin Berry.
\newblock Optimization of the {Lucy} interplanetary trajectory via two-point direct shooting.
\newblock In {\em AAS/AIAA Astrodynamics Specialist Conference}, number AAS 19-663, 2019.

\bibitem{bib:jga1}
Takanao Saiki, Jun Matsumoto, Osamu Mori, and Jun’ichiro Kawaguchi.
\newblock Solar power sail trajectory design for {Jovian} {Trojan} exploration.
\newblock {\em Transactions of the Japan Society for Aeronautical and Space Sciences, Aerospace Technology Japan}, 16(5):353--359, 2018.

\bibitem{bib:jga2}
Yuki Takao, Osamu Mori, Masanori Matsushita, and Ahmed~Kiyoshi Sugihara.
\newblock Solar electric propulsion by a solar power sail for small spacecraft missions to the outer solar system.
\newblock {\em Acta Astronautica}, 181:362--376, 2021.

\bibitem{bib:cassini}
Fernando Peralta and Steve Flanagan.
\newblock {Cassini} interplanetary trajectory design.
\newblock {\em Control Engineering Practice}, 3(11):1603--1610, 1995.

\bibitem{bib:mga1dsm}
Jacob~A Englander, Bruce~A Conway, and Trevor Williams.
\newblock Automated mission planning via evolutionary algorithms.
\newblock {\em Journal of Guidance, Control, and Dynamics}, 35(6):1878--1887, 2012.

\bibitem{bib:trajopt1}
Massimiliano Vasile and Paolo De~Pascale.
\newblock Preliminary design of multiple gravity-assist trajectories.
\newblock {\em Journal of Spacecraft and Rockets}, 43(4):794--805, 2006.

\bibitem{bib:trajopt2}
Dario Izzo, Victor~M Becerra, Darren~R Myatt, Slawomir~J Nasuto, and J~Mark Bishop.
\newblock Search space pruning and global optimisation of multiple gravity assist spacecraft trajectories.
\newblock {\em Journal of Global Optimization}, 38:283--296, 2007.

\bibitem{bib:trajopt3}
Sam Wagner and Bong Wie.
\newblock Hybrid algorithm for multiple gravity-assist and impulsive {Delta-V} maneuvers.
\newblock {\em Journal of Guidance, Control, and Dynamics}, 38(11):2096--2107, 2015.

\bibitem{bib:pso1}
Fernando~Alonso Zotes and Matilde~Santos Penas.
\newblock Particle swarm optimisation of interplanetary trajectories from {Earth} to {Jupiter} and {Saturn}.
\newblock {\em Engineering Applications of Artificial Intelligence}, 25(1):189--199, 2012.

\bibitem{bib:pso2}
Mauro Pontani, Pradipto Ghosh, and Bruce~A Conway.
\newblock Particle swarm optimization of multiple-burn rendezvous trajectories.
\newblock {\em Journal of Guidance, Control, and Dynamics}, 35(4):1192--1207, 2012.

\bibitem{bib:mbh}
CH~Yam, DD~Lorenzo, and D~Izzo.
\newblock Low-thrust trajectory design as a constrained global optimization problem.
\newblock {\em Journal of Aerospace Engineering}, 225(11):1243--1251, 2011.

\bibitem{bib:dvega}
Stefano Campagnola and Ryan~P Russell.
\newblock Endgame problem part 1: {V}-infinity-leveraging technique and the leveraging graph.
\newblock {\em Journal of Guidance, Control, and Dynamics}, 33(2):463--475, 2010.

\bibitem{bib:pagmo}
Francesco Biscani and Dario Izzo.
\newblock A parallel global multiobjective framework for optimization: pagmo.
\newblock {\em Journal of Open Source Software}, 5(53):2338, 2020.

\bibitem{bib:uop}
Damon Landau, Alex Davis, and Reza Karimi.
\newblock Trajectory options for a {Uranus Orbiter and Prob}e.
\newblock In {\em AAS/AIAA Astrodynamics Specialist Conference}, AAS 23-460, Big Sky, MT, USA, 2023.

\bibitem{bib:dcnlp}
Charles~R Hargraves and Stephen~W Paris.
\newblock Direct trajectory optimization using nonlinear programming and collocation.
\newblock {\em Journal of Guidance, Control, and Dynamics}, 10(4):338--342, 1987.

\bibitem{bib:ds1}
Marc~D Rayman, Philip Varghese, David~H Lehman, and Leslie~L Livesay.
\newblock Results from the {Deep Space} 1 technology validation mission.
\newblock {\em Acta Astronautica}, 47(2-9):475--487, 2000.

\bibitem{bib:dawn}
Marc~D Rayman and Keyur~C Patel.
\newblock The {Dawn} project's transition to mission operations: {On} its way to rendezvous with (4) {Vesta} and (1) {Ceres}.
\newblock {\em Acta Astronautica}, 66(1-2):230--238, 2010.

\bibitem{bib:mu10}
Kazutaka Nishiyama, Satoshi Hosoda, Ryudo Tsukizaki, and Hitoshi Kuninaka.
\newblock In-flight operation of the {Hayabusa2} ion engine system on its way to rendezvous with asteroid 162173 {Ryugu}.
\newblock {\em Acta Astronautica}, 166:69--77, 2020.

\bibitem{bib:psyche}
John~S Snyder, Vernon~H Chaplin, Dan~M Goebel, Richard~R Hofer, Alejandro Lopez~Ortega, Ioannis~G Mikellides, Taylor Kerl, Giovanni Lenguito, Faraz Aghazadeh, and Ian Johnson.
\newblock Electric propulsion for the {Psyche} mission: {Development} activities and status.
\newblock In {\em AIAA Propulsion and Energy 2020 Forum}, page 3607, 2020.

\end{thebibliography}

\end{document}